\documentclass[journal,twoside]{IEEEtran}
\usepackage{amsfonts,bm,booktabs,cite,graphicx,hyperref,multirow,textcomp,url}
\usepackage{algorithm}
\usepackage{subfigure}
\usepackage{algpseudocode}
\usepackage{amsmath}
\usepackage{mathrsfs}
\usepackage{float}
\usepackage{color}
\usepackage{booktabs}
\usepackage{bbding}
\usepackage[flushleft]{threeparttable}

\makeatletter 
\renewcommand{\@thesubfigure}{\hskip\subfiglabelskip}
 \makeatother
\begin{document}
\title{Recent Advances of End-to-End Video Coding Technologies for AVS Standard Development}
\author{
	Xihua Sheng,
	Xiongzhuang Liang,
	Chuanbo Tang,
	Zhirui Zuo,
	Yifan Bian,
	Yutao Xie,
	Zhuoyuan Li,
	Yuqi Li, \\
	Hui Xiang,
	Li Li, \IEEEmembership{Senior Member, IEEE}, 
	Dong Liu, \IEEEmembership{Senior Member, IEEE} \\

\thanks{
Date of current version \today.\par 

The authors are with the MOE Key Laboratory of Brain-Inspired Intelligent
Perception and Cognition, University of Science and Technology of China,
Hefei 230093, China (e-mail: xhsheng@mail.ustc.edu.cn; lxz123@mail.ustc.edu.cn; cbtang@mail.ustc.edu.cn;  zuozr@mail.ustc.edu.cn; togelbian@gmail.com; yutaoxie@mail.ustc.edu.cn; zhuoyuanli@mail.ustc.edu.cn;  lyq010303@mail.ustc.edu.cn; xh0603@mail.ustc.edu.cn; lil1@ustc.edu.cn; dongeliu@ustc.edu.cn).\par

Corresponding author: Dong Liu.\par
}
}

\markboth{Submitted to IEEE Transactions on Circuits and Systems for Video Technology}{Recent Advances of End-to-End Video Coding Technologies for AVS Standard Development}

\maketitle
\begin{abstract}
Video coding standards are essential to enable the interoperability and widespread adoption of efficient video compression technologies. In pursuit of greater video compression efficiency, the AVS video coding working group launched the standardization exploration of end-to-end intelligent video coding, establishing the AVS End-to-End  Intelligent Video Coding Exploration Model (AVS-EEM) project. A core design principle of AVS-EEM is its focus on practical deployment, featuring inherently low computational complexity and requiring strict adherence to the common test conditions of conventional video coding. This paper details the development history of AVS-EEM and provides a systematic introduction to its key technical framework, covering model architectures, training strategies, and inference optimizations. These innovations have collectively driven the project's rapid performance evolution, enabling continuous and significant gains under strict complexity constraints. Through over two years of iterative refinement and collaborative effort, the coding performance of AVS-EEM has seen substantial improvement. Experimental results demonstrate that its latest model achieves superior compression efficiency compared to the conventional AVS3 reference software, marking a significant step toward a deployable intelligent video coding standard.

\end{abstract}
\IEEEpeerreviewmaketitle

\section{Introduction}
Video serves as a fundamental form of multimedia content, widely used in applications such as television broadcasting, video streaming, and social media. Particularly with the rise of internet services, an ever-increasing volume of video is being shared and distributed online. Concurrently, the demand for higher visual quality---characterized by increased resolution, frame rate, and dynamic range---continues to grow. This trend leads to a dramatic rise in raw video data volume, making efficient video compression essential for practical storage and transmission.

Video coding is a core technology that enables compact digital representation of video signals by exploiting spatial, temporal, and statistical redundancies. To ensure interoperability and widespread adoption, the development of video coding has been deeply intertwined with standardization efforts led by major international bodies, notably the ITU-T and ISO/IEC. Their collaboration has produced a series of seminal standards, each addressing the pressing demands of its era. The evolution began with H.261~\cite{liou1991overview}, which established the foundational hybrid video coding framework. Its successor, MPEG-2~\cite{okubo1995mpeg}, was instrumental in the global transition to digital television broadcasting. As video consumption shifted to the Internet, H.264/AVC~\cite{wiegand2003overview} emerged as the enabling technology, striking an optimal balance between compression efficiency and computational complexity to become the ubiquitous cornerstone for online video. The subsequent push for higher-fidelity entertainment drove the development of H.265/HEVC~\cite{sullivan2012overview}, which efficiently supports 4K Ultra-High-Definition (UHD) content. The latest standard, H.266/VVC~\cite{bross2021overview}, further advances compression to facilitate 8K UHD, high dynamic range, and immersive media applications. A defining characteristic of this generational progression is the consistent achievement of approximately 50\% bit-rate reduction at equivalent subjective quality, delivering the necessary compression gains to manage exponentially growing video data volumes.\par

Parallelly, the Audio and Video Coding Standard (AVS) working group of China has developed its own lineage of high-performance video coding standards since its establishment in 2002. Although its work started about a decade later than its international counterparts, AVS has rapidly evolved through three generations, each distinguished by hardware-friendly design and a focus on practical deployment. The first generation, AVS1~\cite{fan2004overview}, offered a streamlined alternative to H.264/AVC with notably lower implementation complexity, facilitating its adoption in digital broadcasting. The second generation, AVS2~\cite{ma2015avs2}, not only achieved compression efficiency comparable to H.265/HEVC but also pioneered ``smart" coding features, most notably a background-picture-based mode that significantly enhances compression for surveillance videos by exploiting temporal stationarity. The latest generation, AVS3~\cite{zhang2019recent}, targets 8K UHD, high dynamic range, and smart video applications. It introduces advanced tools such as flexible partitioning, cross-component prediction, and decoder-side motion refinement. Performance evaluations show AVS3 delivers coding efficiency on par with H.266/VVC while maintaining a favorable decoding complexity trade-off. Furthermore, AVS3 has seen rapid industrial adoption, including the world's first 8K decoder chip and its official use for 8K UHD broadcasting by China Central Television, underscoring its role as a mature and deployable next-generation coding standard.\par

The compression efficiency gains from evolving traditional hybrid coding frameworks are becoming increasingly marginal, where the introduction of new coding tools yields only minor compression improvements at the cost of significantly increased design and computational complexity. This has motivated the search for a paradigm beyond incremental engineering. Deep learning offers a promising alternative through end-to-end intelligent compression, where neural networks jointly optimize the entire coding pipeline, opening up new possibilities for compression efficiency. Several international standardization initiatives are actively exploring this new frontier. JPEG-AI~\cite{ascenso2023jpeg,alshina2024jpeg} seeks to create the first intelligent image coding standard, with a unique focus on a single bitstream that serves both high-quality human viewing and efficient machine analysis.
The IEEE 1857.11 working subgroup also develops a intelligent image coding standard, which  gathers several advanced neural network-based coding architectures, such as a wavelet-like transform-based codec~\cite{dong2024wavelet} and a decoupled entropy model-based codec~\cite{zhang2023end}, for  both lossy and lossless image compression. MPEG VCM~\cite{duan2020video,gao2021recent,yang2024video} shifts the focus to machine consumption, pioneering a novel video and visual feature compression paradigm optimized for AI task performance rather than pixel reconstruction. 
Most closely related to the focus of this paper, MPAI-EEV~\cite{jia2023mpai} is developing a fully intelligent video codec, which systematically replaces key components like motion estimation, motion compression, motion compensation, and residual compression with learnable neural networks.\par

AVS began its exploration of end-to-end intelligent video coding in 2023. At the 85-th AVS meeting in June, the first related technical proposal was formally submitted. Following this, the AVS video coding working group officially established the AVS End-to-End Intelligent Video Coding Exploration Model (AVS-EEM) project at the 86-th meeting in August 2023. A defining feature of the AVS-EEM project is its emphasis on practical feasibility. From the beginning, strict complexity constraints were set: encoding complexity was capped at 300 KMAC/pixel and decoding complexity at 200 KMAC/pixel. All evaluations were required to follow the AVS3 common test conditions strictly, ensuring that performance was measured against a consistent and practical benchmark. This focus on complexity control is a key design principle of AVS-EEM, aiming to make end-to-end intelligent video coding viable for real-world deployment.\par

Since the project started, it has received many technical contributions from the global research and industry community. These proposals mainly cover model architectures, training techniques, and inference optimization methods. Development followed an iterative process. The first baseline model, AVS-EEM v0.1, was released on September 8, 2023, which was followed by ongoing development that produced the latest model, AVS-EEM v9.2~\footnote{The training and testing codes of AVS-EEM can be accessed at https://openi.pcl.ac.cn/shengxihua/AVS-EEM once authorized.}, in January 2026. A significant improvement in coding performance is evident across these versions. Under standard low-delay test conditions for YUV420 video, the initial v0.1 model showed a large performance gap compared to the conventional AVS3 benchmark (HPM-15.1), with BD-rate increases of 201.37\% for luma (Y), 140.48\% for chroma U, and 121.13\% for chroma V. In contrast, the latest AVS-EEM v9.2 model achieves BD-rate reductions of --4.14\% for luma, --9.58\% for U, and --24.72\% for V. This progress---from a substantial deficit to clear coding gains---demonstrates the effective collaboration of numerous contributors from academia and industry. It positions AVS-EEM as a practical step toward a standardized end-to-end intelligent video coding solution.

This paper provides a systematic overview of the standard development efforts of AVS end-to-end intelligent video coding. Section~\ref{sec:related_work} first reviews the evolution of conventional AVS video coding standards and related end-to-end intelligent image and video coding standards. Section~\ref{sec:history} details the development history of AVS-EEM. Section IV provides an introduction of the key techniques comprising the AVS-EEM framework, covering model architectures, training strategies, and inference optimizations. Section~\ref{sec:experiments} presents a compression performance comparison against the conventional AVS3 anchor and an analysis of complexity. Section~\ref{sec:future_work} outlines potential future work for the AVS-EEM project. Finally, Section~\ref{sec:conclusion} concludes the paper.

\par
\section{Related Work}\label{sec:related_work}
\subsection{Evolution of AVS Video Coding Standards}
The AVS working group of China, established in 2002, has developed three generations of video coding standards over the past two decades: AVS1~\cite{fan2004overview}, AVS2~\cite{ma2015avs2}, and AVS3~\cite{zhang2019recent}. Each generation has targeted evolving application needs while progressively improving compression efficiency and incorporating hardware-friendly design. AVS1, finalized in 2003 and ratified as a Chinese national standard in 2006, was designed for standard-definition and high-definition digital TV broadcasting. It adopted a hybrid coding framework with tools such as 8$\times$8 integer transform, quarter-pixel motion interpolation with a 4-tap filter, and context-based 2D-VLC entropy coding. AVS1 offered competitive performance to H.264/AVC~\cite{wiegand2003overview} at lower complexity, making it suitable for broadcast and storage applications.
\par
AVS2, standardized in 2016, was developed as a successor to AVS1 with the goal of providing higher compression efficiency for UHD content and specialized applications such as surveillance and videoconferencing. The standard introduced a flexible coding structure based on coding units, prediction units, and transform units, supporting block partitions from 4$\times$4 up to 64$\times$64. AVS2 introduced several advanced tools such as a flexible quadtree-based block structure, asymmetric motion partitions, 33 angular intra prediction modes, and background modeling for surveillance scenarios. It also enhanced in-loop filtering with sample adaptive offset and adaptive loop filter. A distinctive ``smart" feature of AVS2 is its background-picture-based coding mode, which constructs a long-term reference picture through background modeling, significantly improving compression for surveillance videos and facilitating object detection and tracking. AVS2 achieved approximately 30\% bit-rate reduction over AVS1 and performed comparably to H.265/HEVC~\cite{sullivan2012overview}. \par
The latest generation, AVS3, was developed in two phases: Main Profile (finalized in 2019) and High Profile (finalized in 2021). AVS3 focuses on 8K UHD, high dynamic range, and smart video applications. It introduced key technical innovations such as extended block partitioning, cross-component prediction, affine motion compensation, and decoder-side motion refinement. Performance evaluations show that AVS3 Main Profile achieves about 23.5\% bit-rate saving over AVS2 under random access configuration, while AVS3 High Profile reaches up to 34\% reduction. The standard also maintains a favorable complexity-performance trade-off, with decoding complexity well controlled relative to H.266/VVC~\cite{bross2021overview}. \par
The evolution of AVS reflects a consistent trend toward higher coding efficiency and the greater adaptability to emerging content types, laying a solid foundation for the ongoing standardization of end-to-end intelligent video coding.
\begin{figure*}[t]
  \centering
   \includegraphics[width=0.9\linewidth]{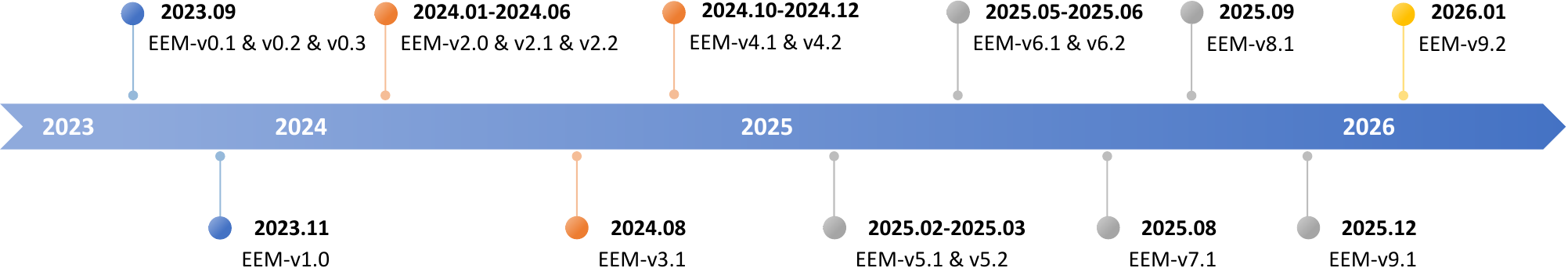}
      \caption{Development history of  AVS end-to-end intelligent video coding exploration model (AVS-EEM).}
   \label{fig:timeline}
\end{figure*}
\subsection{Standardization Efforts of End-to-End Intelligent Image and Video Coding}
JPEG-AI~\cite{ascenso2023jpeg,alshina2024jpeg} is an ongoing international standardization effort initiated by the JPEG Committee in 2019, aiming to develop the first end-to-end intelligent image coding standard. Its main objective is to define a compact, single-stream compressed representation that simultaneously supports high-quality image reconstruction for human viewing and efficient feature reuse for machine-vision tasks---such as classification, object detection, and segmentation---without requiring full decoding. This standard adopts a deep neural network-based coding framework, featuring a multi-branch codec architecture that offers two encoder variants (with or without attention transformers) and three decoder variants of increasing complexity. This design allows devices to select the appropriate performance-complexity trade-off according to their capabilities.  JPEG-AI also supports progressive decoding, where reconstruction quality improves as more latent channels are decoded, as well as partial-picture decoding through tiled latent representations, enabling region-of-interest reconstruction and random spatial access. In terms of compression performance, against H.266/VVC Intra, its simplest configuration achieves about 12\% BD-rate saving with encoding speeds up to 2000$\times$ faster and GPU decoding 2$\times$ faster. Its highest configuration reaches 27\% BD-rate saving while maintaining faster encoding and GPU-accelerated decoding. \par

The IEEE 1857.11 working subgroup, also known as the Future Video Coding Study Group, was established in 2012 to develop advanced neural-network-based image coding standards. In response to its Call for Proposals (CfP), several notable learning-based compression frameworks have been contributed. Dong et al.~\cite{dong2024wavelet} proposed iWaveV3, an end-to-end wavelet-inspired image codec that supports both lossy and lossless image compression within a unified model. The framework incorporates an affine wavelet-like transform, a perception-aware quality metric, and advanced training techniques such as soft-then-hard quantization and online optimization. On the Kodak dataset, iWaveV3 achieves an 11.07\% BD-rate gain over H.266/VVC Intra while preserving competitive perceptual quality. Zhang et al.~\cite{zhang2023end} presented an end-to-end image codec with a decoupled architecture, which separates entropy decoding from latent sample reconstruction to enable parallel processing. Coupled with wavefront processing, adaptive quantization, and device-interoperable quantization strategies, the codec achieves a 29.6\% average BD-rate reduction against H.266/VVC on the JPEG AI CfP test set while providing 2.44$\times$ faster GPU-accelerated decoding. Its design also supports tiling for memory-constrained devices and variable-rate coding via resampling or quantization adjustment, making it highly suitable for real-world applications.\par

The Moving Picture Experts Group (MPEG) initiated the Video Coding for Machines (VCM)~\cite{duan2020video,gao2021recent,yang2024video} standardization activity in July 2019. Its primary goal is to develop efficient compression methods for video and visual features that are optimized for machine vision tasks, while optionally supporting hybrid human-machine consumption. MPEG VCM organizes its standardization efforts into two complementary tracks. Track 1: Feature Coding serves as the main pathway for end-to-end optimization. It focuses on compressing intermediate feature maps extracted by a neural network at the edge device. These compressed features are transmitted to a server, where a downstream network completes the target vision task. This pipeline embodies a fully learnable, end-to-end paradigm, allowing joint optimization of the entire chain---feature extraction, compression, transmission, and task inference---for both bit-rate efficiency and task accuracy. In contrast, Track 2: Image and Video Coding primarily investigates enhancements to conventional video codecs (e.g., H.266/VVC) or region-of-interest-based preprocessing to improve machine-vision performance from reconstructed pixels. Although some proposals under this track incorporate deep-learning modules, the overall architecture largely remains within a classical hybrid coding framework, rather than constituting a fully end-to-end learned compression pipeline.\par

The Moving Picture, Audio, and Data Coding by Artificial Intelligence (MPAI) organization launched the MPAI-EEV (End-to-End Video Coding) project~\cite{jia2023mpai} in December 2021. This initiative aims to develop a video coding standard that operates entirely outside the conventional hybrid coding framework by leveraging data-trained neural networks to jointly optimize all compression modules in an end-to-end manner. 
The technical roadmap of MPAI-EEV follows a progressive, model-driven approach. Beginning with the publicly available OpenDVC framework (designated as EEV-0.1), the project has iteratively released four verification models (EEV-0.1 to EEV-0.4). EEV-0.2 integrates an enhanced motion-compensation network with residual channel attention to improve inter-prediction quality. EEV-0.3 further introduces a coarse-to-fine residual modeling module and an in-loop restoration network to reconstruct fine-grained textures and suppress artifacts. EEV-0.4 adopts a motion-decoupling scheme that decomposes motion representation into intrinsic and compensatory components using stacked ConvLSTM units, enabling more efficient temporal context modeling. 
Performance evaluation under a slightly different condition to the HEVC/VVC common test conditions---using a low-delay configuration with an intra period of 16 and RGB-PSNR as the distortion metric---shows that EEV-0.4 achieves a significant BD-rate reduction of --39.72\% over HEVC on drone-captured (UAV) sequences, and --18.70\% on standard HEVC test sequences and the UVG dataset. It should be noted, however, that the computational complexity of EEV-0.4 remains high, measured at approximately 3127 KMAC/pixel.

\section{Development History of AVS-EEM}\label{sec:history}
The initiative of  AVS end-to-end intelligent video coding project began at the 85-th AVS meeting in June 2023. At this meeting, an intelligent video coding framework based on temporal context mining~\cite{sheng2022temporal,sheng2024spatial} was first introduced. Initial results indicated that the framework could surpass the compression efficiency of the AVS3 reference software HPM-15.1 under low-delay configuration, although at a high computational cost exceeding 1800 KMAC/pixel. Consequently, at the 86-th meeting in August 2023, the AVS video coding working group formally established the AVS-EEM project. Critical development constraints were defined, cencoding complexity at 300 KMAC/pixel and decoding complexity at 200 KMAC/pixel, while requiring strict adherence to the AVS3 common test conditions. As illustrated in Fig.~\ref{fig:timeline}, the AVS-EEM project has since progressed through a series of rapid and structured iterations. The first version, AVS-EEM v0.1, was released on September 8, 2023. It was followed by numerous subsequent releases throughout 2024 and 2026 (e.g., v1.0 in November 2023, v2.0 in January 2024, up to v9.2 in early 2026), reflecting continuous refinement of the framework.
\par
A significant evolution in coding performance is evident across these versions. When encoding YUV420 video under standard test conditions, the initial v0.1 exhibited a substantial BD-rate increase over HPM-15.1---201.37\% for luma (Y), and 140.48\% and 121.13\% for the chroma components (U and V). In contrast, the latest model, AVS-EEM v9.2, achieves BD-rate reductions of --4.14\% in luma and --9.58\% and --24.72\% in chroma. This remarkable progress, transitioning from a significant performance gap to measurable coding gains, results from the collaborative efforts of numerous researchers from academia and industry. Their contributions have introduced a wide array of efficient model architectures, advanced training strategies, and inference optimizations into the framework. As a result, substantial compression performance improvements have been achieved while rigorously maintaining the prescribed computational complexity limits.
\begin{figure*}[t]
  \centering
   \includegraphics[width=0.95\linewidth]{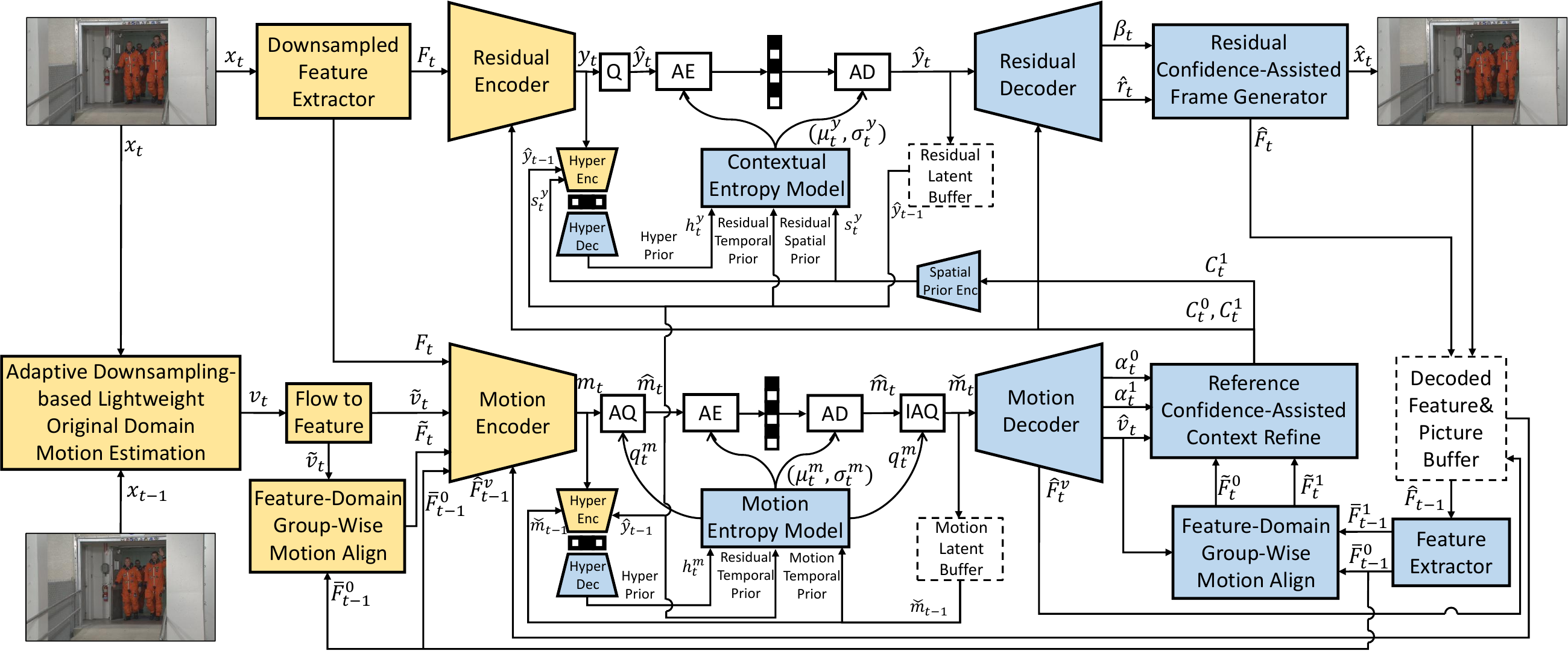}
      \caption{Architecture of the latest AVS end-to-end intelligent video coding exploration model v9.2.}
   \label{fig:framework}
\end{figure*}
\section{Key Techniques of AVS-EEM}
\subsection{Overview}
The AVS-EEM framework follows a conditional coding paradigm~\cite{li2021deep,sheng2022temporal}, maintaining the essential motion-residual architecture of traditional hybrid video coding while implementing all its components as learnable neural networks.  As illustrated in Fig.~\ref{fig:framework}, its architecture comprises two core branches: a motion branch and a residual branch. Within the motion branch, a motion estimation module first derives the inter-frame motion field $v_t$. This motion field is compressed for transmission using a motion encoder and an associated motion entropy model, and subsequently decompressed by a motion decoder into $\hat{v}_t$. The decoded motion field drives a motion compensation module to predict temporal contexts of the current frame $x_t$. In the residual branch, conditioned on the predicted temporal contexts, the current frame $x_t$ is compressed through a residual encoder and a residual entropy model, and then reconstructed by a residual decoder and a frame generator to produce the reconstructed frame $\hat{x}_t$. The entire framework is optimized in an end-to-end manner using a rate-distortion loss function, which jointly balances bit rate and reconstruction distortion. The following sections detail the key coding techniques introduced within this framework to advance compression performance under strict computational complexity constraints.
\begin{figure}[t]
  \centering
   \includegraphics[width=0.7\linewidth]{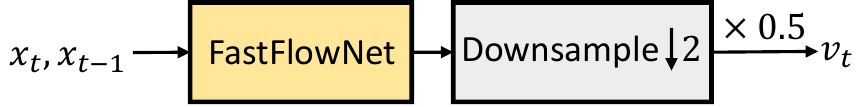}
      \caption{Original domain downsampled motion estimation.}
   \label{fig:motion_estimation}
\end{figure}

\subsection{Key Model Architectures of AVS-EEM}
\subsubsection{Original Domain Downsampled Motion Estimation}\label{motionestimation}
AVS-EEM v1.0 proposes an original domain downsampled motion estimation. As shown in Fig.~\ref{fig:motion_estimation}, the network operates directly in the original pixel domain---using the current frame $x_t$ and the original reference frame $x_{t-1}$---rather than the reconstructed frame $\hat{x}_{t-1}$. This design prevents accuracy degradation from compression artifacts introduced in prior frames. To further reduce computational overhead, the estimated motion field $v_t$ is spatially downsampled by a factor of two before being processed by the motion encoder. Correspondingly, the motion decoder reconstructs a half-resolution motion field $\hat{v}_t$. This downsampling strategy significantly lowers the complexity of motion compression while retaining sufficient precision for effective motion compensation.

\begin{figure}[t]
  \centering
   \includegraphics[width=\linewidth]{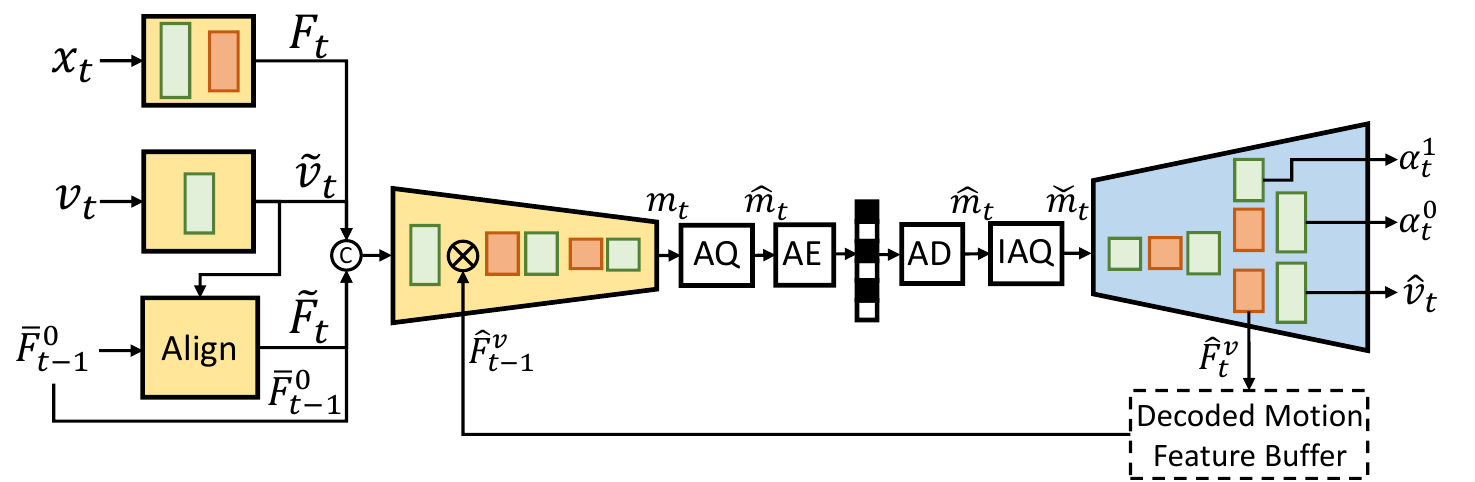}
      \caption{Content and motion feature-assisted motion compression.}
   \label{fig:motion_compression}
\end{figure}

\subsubsection{Content and Motion Feature-Assisted Motion Compression}
AVS-EEM v1.0 proposes to perform motion compression in the feature domain. As illustrated in Fig.~\ref{fig:motion_compression}, the two-channel optical flow $v_t$ is first transformed by a convolutional layer into a motion feature $\tilde{v}_t$. A motion encoder, composed of stride-2 convolutional layers and residual blocks, compresses  $\tilde{v}_t$ into a compact latent representation $m_t$. After adaptive quantization (AQ, Section~\ref{sec:motion_entropy}), the quantized motion latent representation $\hat{m}_t$ are entropy-coded by an arithmetic encoder according to the probability distribution estimated from a motion entropy model (Section~\ref{sec:motion_entropy}). On the decoder side, the bitstream is decoded to  $\hat{m}_t$, which is then inversely adaptive quantized (IAQ) to obtain $\check{m}_t$. A motion decoder---comprising sub-pixel layers, residual blocks, and convolutional layers---reconstructs the motion feature $\hat{v}_t$ from $\check{m}_t$ and two reference confidence coefficients $\alpha_{t}^0$, $\alpha_{t}^1$ (Section~\ref{sec:tcm}). To exploit the correlation between video content and motion for higher compression efficiency, the motion encoder is additionally conditioned on several auxiliary features: (1) a downsampled content feature $F_t$ of the current frame; (2) a derived feature $\bar{F}_{t-1}^0$ extracted from the reference feature $\hat{F}_{t-1}$ (Section~\ref{sec:tcm}); and (3) an aligned feature $\tilde{F}_t$, obtained by applying feature-domain group-wise alignment (Section~\ref{sec:align}) to $\bar{F}_{t-1}^0$ using the motion feature $\tilde{v}_t$. Furthermore, to leverage temporal motion coherence, an intermediate feature from the previous frame's motion decoder (at $\frac{1}{4}$ resolution), denoted as $\hat{F}_{t-1}^v$, is also provided as a condition to the motion encoder. Incorporating these content and motion-related features significantly improves the efficiency of motion compression.
\begin{figure}[t]
  \centering
   \includegraphics[width=\linewidth]{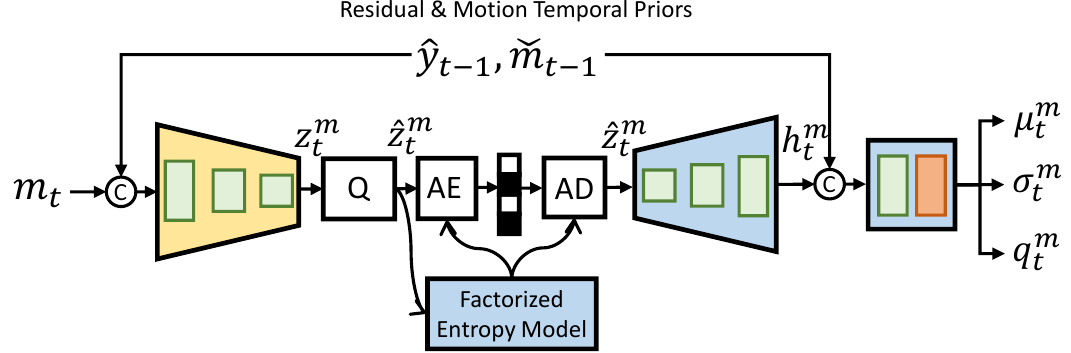}
      \caption{Residual and motion temporal prior-assisted motion entropy model.}
   \label{fig:motion_entropy}
\end{figure}

\subsubsection{Residual and Motion Temporal Prior-Assisted Motion Entropy Modeling}\label{sec:motion_entropy}
AVS-EEM v0.1 adopts a hyper-prior entropy model to estimate the probability distribution of the motion latent representation $\hat{m}_t$. As presented in Fig.~\ref{fig:motion_entropy}, the model first compresses $m_t$ via a hyper-encoder composed of stride-2 convolutional layers, producing a more compact hyper-latent representation  $z_t^m$. After quantization, $z_t^m$ is entropy-coded by an arithmetic encoder according to a factorized entropy model. At the decoder, the bitstream is decoded to $z_t^m$, which is then passed through a hyper-decoder built with sub-pixel layers to recover the hyper-prior $h_t^m$. To improve the compression efficiency, AVS-EEM v1.0 proposes to feed the the motion latent representation $\check{m}_{t-1}$ into the hyper-encoder and AVS-EEM v4.1 further proposes to feed the residual latent representation $\hat{y}_{t-1}$. Simultaneously, these priors are fused with the hyper-prior $h_t^m$ and processed by a gather layer consisting of convolutional and residual blocks to produce the mean $\mu_{t}^m$ and variance $\sigma_{t}^m$ of the estimated Gaussian distribution for $\hat{m}_t$. Additionally,  it outputs a quantization step $q_t^m$, which is employed in the adaptive (and inverse) quantization. $q_t^m$ scales the motion latent representation $m_t$ before the round operator during encoding, and inversely scales the entropy-decoded quantization latent representation $\hat{m}_t$ during decoding.
\begin{figure}[t]
  \centering
   \includegraphics[width=\linewidth]{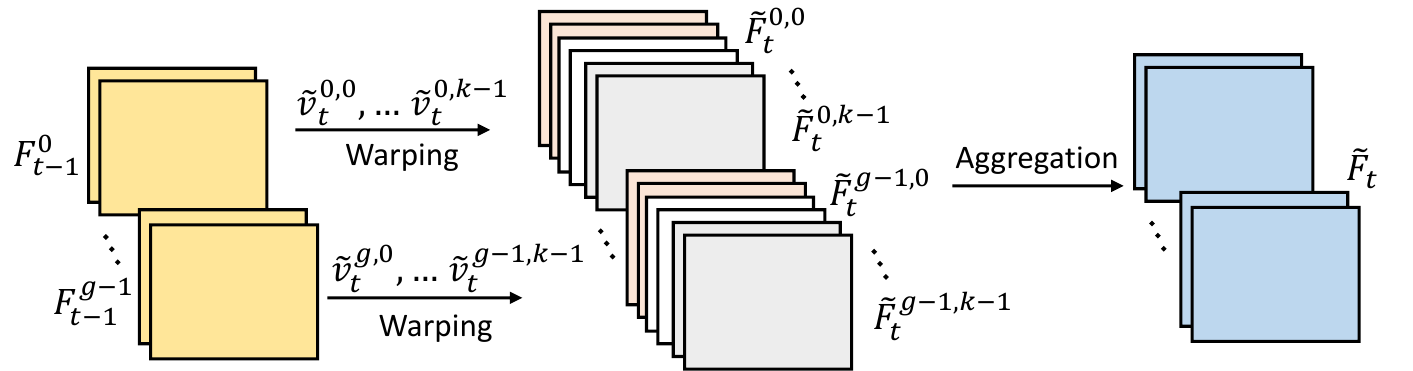}
      \caption{Feature-domain group-wise motion alignment.}
   \label{fig:motion_compensation}
\end{figure}

\subsubsection{Feature-Domain Group-Wise Motion Alignment}\label{sec:align}
To mitigate spatial misalignment between reference features and the current frame, thereby obtaining more accurate inter-frame prediction, AVS-EEM 2.0 proposes a feature-domain group-wise motion alignment method. As shown in Fig.~\ref{fig:motion_compensation}, given a reference feature  $F_{t-1}$, it is first split along the channel dimension into $g$ groups: $F_{t-1}^0, F_{t-1}^1, ..., F_{t-1}^{g-1}$. Similarly, the motion feature $\tilde{v}_t$, is divided into $g \times k$ two-channel optical flow fields. For the $i$-th group ($i$ = $0, \cdots, g-1$), the corresponding subset of flow fields $\tilde{v}_t^{i,0}, \tilde{v}_t^{i,1}, \cdots, \tilde{v}_t^{i,k-1}$ is used to warp the group feature $F_{t-1}^{i}$ separately, yielding $k$ warped features $\tilde{F}_t^{i,0}, \tilde{F}_t^{i,1}, \cdots, \tilde{F}_t^{i,k-1}$. Finally, all $g \times k$ warped features are aggregated through a $1 \times 1$ convolutional layer to produce the aligned feature $\tilde{F}_t$. This group-wise design increases alignment flexibility and helps capture complex motion patterns more effectively than a monolithic warping operation. 
\begin{figure}[t]
  \centering
   \includegraphics[width=0.85\linewidth]{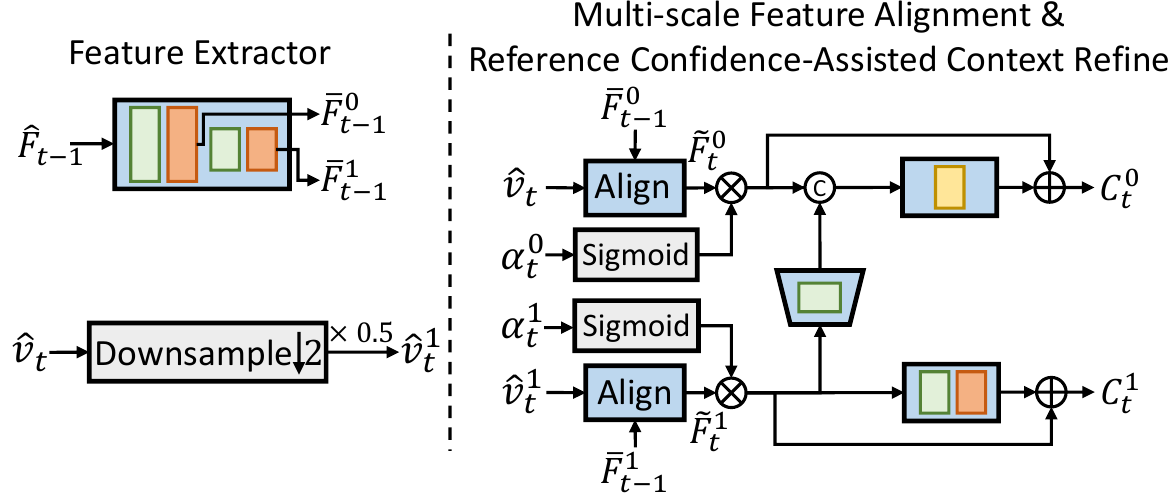}
      \caption{Reference confidence-assisted temporal context mining.}
   \label{fig:temporal_context_mining}
\end{figure}

\subsubsection{Reference Confidence-Assisted Temporal Context Mining}\label{sec:tcm}
AVS-EEM v2.0 proposes a multi-scale temporal context mining method to predict multi-scale temporal contexts for residual coding. The method comprises three key stages: multi-scale feature extraction, multi-scale feature alignment, and context refinement. First, a feature extractor composed of convolutional layers and residual blocks extracts multi-scale features $\bar{F}_{t-1}^0$ ($\frac{1}{2}$ resolution) and $\bar{F}_{t-1}^1$ ($\frac{1}{4}$ resolution) from the propagated reference feature $\hat{F}_{t-1}$. Simultaneously, the reconstructed motion feature $\hat{v}_{t}$  is bilinearly downsampled to obtain $\hat{v}_t^1$ at $\frac{1}{4}$ resolution. Using $\hat{v}_{t}$ and $\hat{v}_t^1$ respectively, feature-domain group-wise motion alignment (Section~\ref{sec:align}) is applied to $\bar{F}_{t-1}^0$ and $\bar{F}_{t-1}^1$, yielding aligned multi-scale predicted features $\tilde{F}_{t-1}^0$ and $\tilde{F}_{t-1}^1$. To adaptively modulate these aligned features according to their reliability, AVS-EEM v5.1 proposes two reference confidence coefficients $\alpha_t^0$ and $\alpha_t^1$ derived from the motion decoder, whose channel and spatial dimensions match $\tilde{F}_{t-1}^0$ and $\tilde{F}_{t-1}^1$, respectively. Element-wise multiplication is performed between the confidence coefficients and the aligned features as shown in Fig.~\ref{fig:temporal_context_mining}. This weighting emphasizes regions and channels where the motion-aligned prediction is more reliable. The modulated features are then refined through a context-refinement process similar to that described in~\cite{sheng2022temporal}. Finally, the refined multi-scale temporal contexts $C_t^0$ and $C_t^1$ are obtained. To reduce computational cost without sacrificing performance, the convolutional layer and residual block in the larger-scale branch are replaced with a lightweight inception block~\cite{szegedy2017inception} in AVS-EEM v5.1, which further improve efficiency and representation capacity.

\subsubsection{Multi-Scale Temporal Context-Assisted Feature-Domain Residual Compression}
\begin{figure}[t]
  \centering
   \includegraphics[width=\linewidth]{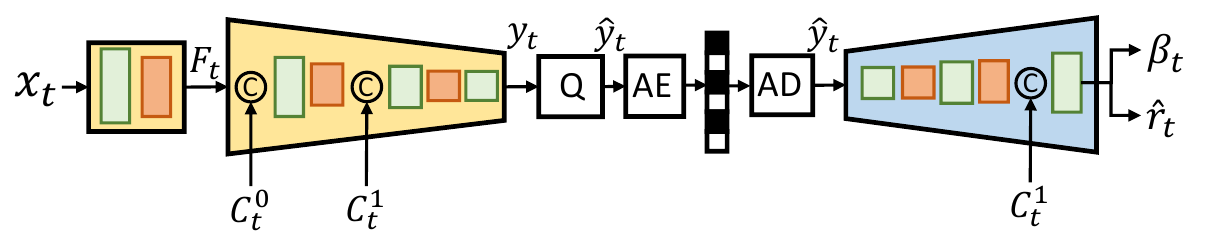}
      \caption{Multi-scale temporal context-assisted feature-domain residual compression.}
   \label{fig:residual_compression}
\end{figure}

AVS-EEM v1.0 proposes to conduct residual compression in the feature domain. As illustrated in Fig.~\ref{fig:residual_compression}, the pixel-domain current frame $x_t$  is first processed by a feature extractor composed of a stride-2 convolutional layer and a residual block, yielding a half-resolution feature $F_t$. A residual encoder, also built with stride-2 convolutional layers and residual blocks, then compresses $F_t$ into a compact latent representation $y_t$. Throughout the encoding process, AVS-EEM v2.0 proposes to inject the predicted multi-scale temporal contexts $C_t^0$ and $C_t^1$ into the encoder to effectively reduce temporal redundancy. After quantization, the quantized latent representation $\hat{y}_t$ is entropy-coded by an arithmetic encoder according to a probability distribution supplied by a residual entropy model (Section~\ref{sec:residual_entropy}). At the decoder, the bitstream is decoded back to $\hat{y}_t$. A residual decoder---consisting of sub-pixel layers and residual blocks---reconstructs the residual feature $\hat{r}_t$ from $\hat{y}_t$ and a residual confidence coefficient $\beta_{t}$  (Section~\ref{sec:frame_generator}). During decoding, the temporal context $C_t^1$ is also fed into the residual decoder to provide complementary temporal information, ensuring high-fidelity reconstruction.
\begin{figure}[t]
  \centering
   \includegraphics[width=0.55\linewidth]{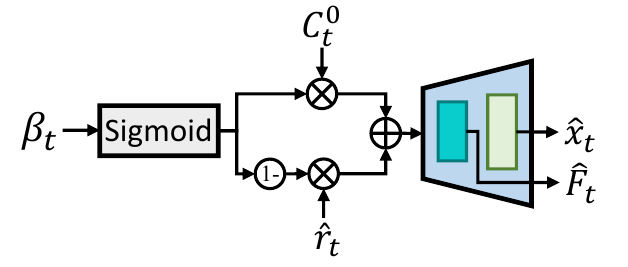}
      \caption{Residual confidence-assisted frame generator.}
   \label{fig:frame_generator}
\end{figure}
\subsubsection{Residual Confidence-Assisted Frame Generator}\label{sec:frame_generator}
After obtaining the residual feature $\hat{r}_t$ and the residual confidence coefficient $\beta_{t}$  from the residual decoder, AVS-EEM v2.0 introduces a residual confidence-assisted frame generator. As shown in Fig.~\ref{fig:frame_generator}, rather than simply concatenating $\hat{r}_t$ and the predicted temporal context $C_t^0$ along the channel dimension, the frame generator modulates the reliability of the temporal context using $\beta_{t}$. For regions where the prediction is accurate, the confidence coefficient approaches 1, while for regions with low prediction accuracy, it approaches 0. As illustrated in Fig. 9, element-wise multiplication and addition are applied between $\hat{r}_t$ and  $C_t^0$, guided by $\beta_{t}$, enabling an adaptive fusion of the residual information with the temporal prediction. The fused feature is then passed to a frame generator built with a U-Net architecture~\cite{li2022hybrid} (introduced in AVS-EEM v5.1) and a sub-pixel layer, producing the final reconstructed frame  $\hat{x}_t$. Simultaneously, a half-resolution reconstructed feature $\hat{F}_t$ from the U-Net output is propagated as the reference feature for the subsequent frame.
\begin{figure}[t]
  \centering
   \includegraphics[width=\linewidth]{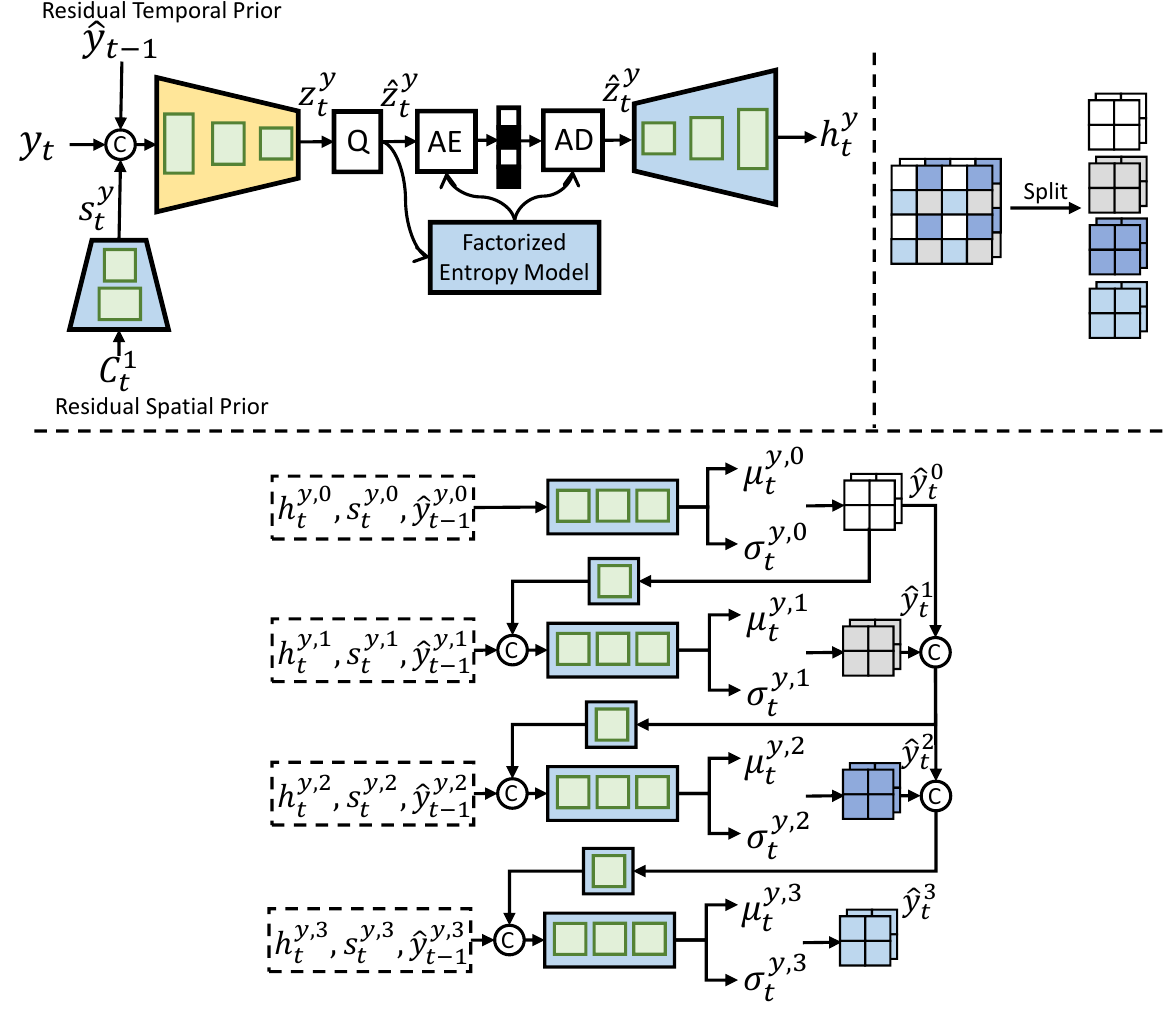}
      \caption{Residual Temporal and Spatial Prior-Assisted Residual Checkerboard Autoregressive Entropy Model.}
   \label{fig:residual_entropy}
\end{figure}

\subsubsection{Residual Temporal and Spatial Prior-Assisted Residual Checkerboard Autoregressive Entropy Modeling}\label{sec:residual_entropy}
AVS-EEM v7.1 proposes a checkerboard-partitioned autoregressive entropy model to estimate the probability distribution of the residual latent representation $\hat{y}_t$. As shown in Fig.~\ref{fig:residual_entropy}, the model first compresses $y_t$ through a hyper-encoder composed of stride-2 convolutional layers, producing a compact hyper-latent representation $z_t^y$. After quantization, $z_t^y$ is entropy-coded by an arithmetic encoder according to a factorized entropy model. At the decoder, the bitstream is decoded to 
$z_t^y$, which is then passed through a hyper-decoder built with sub-pixel layers to generate a hyper-prior $h_t^y$. To improve compression efficiency, AVS-EEM adopts the previously decoded residual latent representation $\hat{y}_{t-1}$ as a temporal prior and a feature $s_t^y$ extracted from the temporal context $C_t^1$ via a spatial-prior encoder as a spatial prior. Both priors are fed as additional conditions into the hyper-encoder. During decoding, the hyper-prior $h_t^y$,  the temporal prior $\hat{y}_{t-1}$, the spatial prior $s_t^y$ and the residual latent $\hat{y}_t$ are each split into four segments following a checkerboard grouping scheme. When processing the $i$-th segment $\hat{y}_t^i$, the corresponding segments of the priors---$h_t^{y,i}$, $\hat{y}_{t-1}^i$, and $s_t^{y,i}$---are concatenated with all previously coded segments $\hat{y}_t^{\leq i}$. These combined priors serve as conditions for estimating the mean $\mu_t^{y,i}$ and variance $\sigma_t^{y,i}$ of a Gaussian distribution for $\hat{y}_t^i$. This segment-wise autoregressive conditioning mechanism enhances probability estimation accuracy while maintaining a fast decoding speed.

\subsubsection{Feature Scaling for Hierarchical Quality}\label{sec:scaling}
To achieve hierarchical quality across frames, AVS-EEM v3.1 proposes a feature scaling method based on learnable, frame-dependent parameters. A set of scaling parameters with dimensions $[R,C,1,1]$ is defined, where $R$ corresponds to the number of quality levels in the hierarchical structure. During training, each frame is associated with a specific quality weight (Section~\ref{sec:hierarchical_training}), and the corresponding scaling parameters are selected accordingly. During inference, parameters are chosen based on a predefined hierarchical structure according to the frame index. The selected scaling parameters are applied via element-wise multiplication to the features of the motion encoder, the motion decoder, the motion temporal prior, the residual temporal prior, the residual spatial prior, the residual encoder, and the residual decoder. This gated feature modulation enables the model to adapt its representation capacity to different quality levels within a sequence while maintaining a single unified architecture.

\subsection{Key Training Techniques of AVS-EEM}
\subsubsection{Progressive Training}
AVS-EEM v0.1 adopts a multi-stage progressive training strategy to ensure stable convergence and effective joint optimization of its modular components. In the first stage, the training focuses exclusively on the modules responsible for generating inter-frame prediction, such as the motion encoder-decoder, the motion entropy model, and an auxiliary predicted-frame generator that transforms the temporal context $C_t^0$ into a pixel-domain predicted frame $\tilde{x}_t$. The distortion $D_t^m$ between the current frame $x_t$ and the predicted frame $\tilde{x}_t$ is measured, and the bit rate $R_t^m$ required to compress the motion latent representation $\hat{m}_t$ along with its hyper-latent $\hat{z}_t^m$ is calculated:
\begin{equation}
L_1 = \lambda \times  D_t^m + R_t^m,
\end{equation}
where $\lambda$ is a Lagrangian multiplier that balances the rate–distortion trade-off. \par
In the second stage, the training objective shifts to optimizing the modules responsible for residual compression and final reconstruction, including the temporal context mining module, the residual encoder-decoder, and the frame generator. The distortion $D_t^r$ between the current frame $x_t$ and the reconstructed frame $\hat{x}_t$ is evaluated, and the bit rate $R_t^r$ for compressing the residual latent representation $\hat{y}_t$ together  with its hyper-latent $\hat{z}_t^y$ is calculated:
\begin{equation}
L_2 = \lambda \times  D_t^r + R_t^r.
\end{equation}
\par
In the third stage, all modules are jointly optimized in an end-to-end manner. The total bit rate  $R_t^r$ is computed as the sum of the bit rates for the motion and residual components:
\begin{equation}
L_3 = \lambda \times  D_t^r + R_t^m + R_t^r.
\label{eq3}
\end{equation}
\par
\begin{figure}[t]
  \centering
   \includegraphics[width=0.8\linewidth]{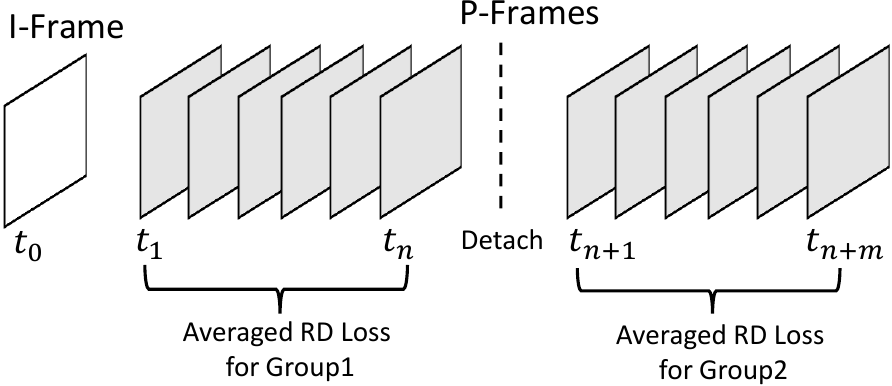}
      \caption{Multi-frame cascaded training.}
   \label{fig:cascaded_training}
\end{figure}

\subsubsection{Hierarchical Quality-Based Training}\label{sec:hierarchical_training}
To enhance rate-distortion performance across frames, AVS-EEM v1.0 proposes a hierarchical quality-based training strategy. This strategy assigns different quality weight to frames according to a predefined hierarchical quality structure. A cyclic quality weight $w_{t\%4} \in \{2.0, 0.2, 0.4, 0.2\}$ is introduced into the loss function, scaling the distortion term according to the frame's position in the hierarchical quality structure:
\begin{equation}
L_4 = w_{t\%4} \times \lambda \times  D_t^r + R_t^m + R_t^r.
\label{eq4}
\end{equation}
where a larger weight prioritizes reconstruction fidelity for key frames.
The training is conducted in a dedicated stage where only the learnable feature-scaling parameters (Section~\ref{sec:scaling}) are updated using Eq.~\eqref{eq4}, while all other network components remain fixed. In subsequent stages, the entire model---including both the feature-scaling parameters and all compression modules---is jointly fine-tuned in an end-to-end manner. This progressive training method stabilizes optimization and ensures that the learned feature scaling effectively supports hierarchical quality across the sequence.
\par 
\begin{figure}[t]
  \centering
   \includegraphics[width=0.7\linewidth]{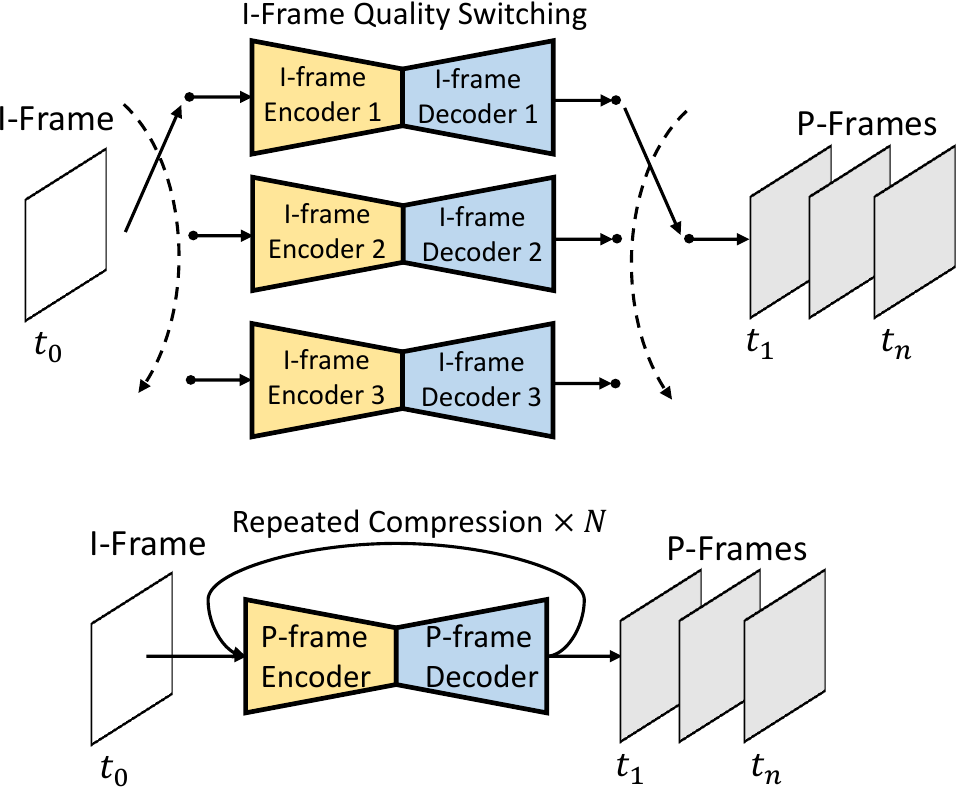}
      \caption{Reference quality-aware training.}
   \label{fig:repeat_compression}
\end{figure}
\subsubsection{Multi-Frame Cascaded Training}
To mitigate reconstruction error accumulation across frames, AVS-EEM v0.1 adopts a multi-frame cascaded training strategy. As shown in Fig.~\ref{fig:cascaded_training}, unlike the per-frame loss functions, where gradients are computed and applied independently for each frame, this strategy averages the loss over multiple consecutive frames before updating the model parameters, enabling joint optimization across a temporal window:
\begin{equation}
L_5=\frac{1}{T} \sum_{t} \{w_{t\%4} \times \lambda \times D_t^r + R_t^m + R_t^r\},
\label{eq5}
\end{equation}
where T is the length of temporal window. 
AVS-EEM employs cascaded training with progressively longer sequences: 6, 15, 20, and 28 P-frames, respectively. To alleviate the increased GPU memory demand that arises with longer sequences, AVS-EEM v5.2 proposes a group-wise cascaded training strategy~\cite{jiang2025ecvc}. For instance, a 28-frame sequence can be split into two subgroups (e.g., 16 and 12 frames). A gradient-detach operation is inserted between the subgroups, preventing gradient flow from the later group to the earlier one, thereby substantially reducing memory overhead while preserving the benefits of multi-frame joint training.
\begin{figure}[t]
  \centering
   \includegraphics[width=0.9\linewidth]{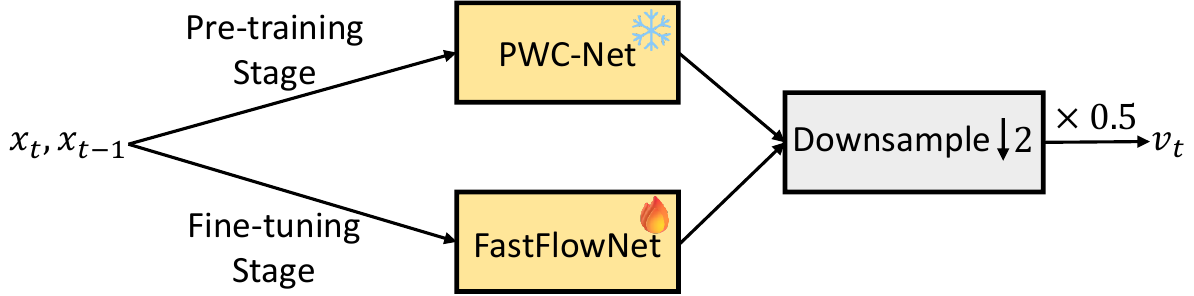}
      \caption{Heavy to light optical flow estimation network training.}
   \label{fig:motion_estimation_training}
\end{figure}

\subsubsection{Reference Quality-Aware Training}
To expose the model to a wider range of reference frame qualities and alleviate the domain gap between training and inference---which otherwise leads to reconstruction-error accumulation---AVS-EEM introduces two complementary reference quality-aware training strategies. The first is an I-frame quality switching strategy proposed in AVS-EEM v7.1. During training, multiple I-frame codecs pre-trained at different $\lambda$ are maintained. For each training iteration, one of these models is randomly selected to compress the original I-frame, generating a reconstructed I-frame with quality corresponding to the target $\lambda$.  This deliberate variation forces subsequent P-frame coding modules to encounter and adapt to reference frames of diverse fidelity levels, thereby improving robustness to quality fluctuations in real-world coding scenarios.\par
The second is a repeated compression-based training strategy proposed in AVS-EEM v1.0. As illustrated in Fig.~\ref{fig:repeat_compression}, each I-frame is iteratively compressed by the P-frame codec for $N$ times, where $N$ is randomly sampled from the range $[1,40]$. This process simulates the various levels of reference-frame degradation that may occur during inference, thereby enabling the model to learn robust representations that are less sensitive to error propagation. As a result, the model becomes more adaptive to low-quality references, ultimately achieving the target reconstruction quality despite the accumulation of compression artifacts over multiple frames.

\subsubsection{Heavy to Light Optical Flow Estimation Network Training}
\label{motion_estimation_training}
AVS-EEM v1.0 proposes a progressive ``heavy-to-light" training strategy for the optical flow estimation network. In the initial phase, a heavyweight pre-trained PWC-Net~\cite{sun2018pwc} (196K MAC/pixel) provides high-accuracy optical flow during training, while a lightweight pre-trained FastFlowNet (27K MAC/pixel) is used for inference, with the weights of PWC-Net kept frozen throughout end-to-end optimization. The second phase refined this strategy by using PWC-Net only in the pre-training stage, then switching to the frozen FastFlowNet for fine-tuning. The most recent phase further enhances this scheme. As shown in Fig.~\ref{fig:motion_estimation_training}, the pre-training stage still relies on frozen PWC-Net, the lightweight FastFlowNet is now integrated into the end-to-end fine-tuning loop, allowing its parameters to be jointly optimized with the rest of the compression pipeline. This ``heavy-to-light" progression effectively balances motion estimation accuracy with practical complexity constraints. 

\subsubsection{Validation During Training}
During the fine-tuning stages, AV-EEM v4.1 proposes an validation strategy to select the optimal model checkpoint. The model is evaluated on a validation set consisting of eight 96-frame 1080p sequences from the USTC-TD~\cite{li2025ustc} dataset. The evaluation criterion is the rate-distortion loss defined in Eq.~\eqref{eq5}. By regularly monitoring this metric across different training stages and epochs, the checkpoint that achieves the best rate-distortion trade-off is retained for final deployment. This validation strategy ensures that the selected model generalizes well and maintains stable performance throughout the iterative training process.

\subsection{Key Inference Optimization Techniques of AVS-EEM}
\subsubsection{Adaptive Downsampling-based Motion Estimation}
A domain gap often exists between the motion characteristics of training sequences and those encountered in test videos, particularly when test content exhibits larger motion amplitudes.  As indicated in~\cite{sheng2024spatial}, spatial downsampling can effectively reduce the magnitude of inter-frame motion. For videos with large motion, if downsampling leads to a smaller frame difference, it indicates that performing motion estimation at the reduced resolution may yield a more accurate prediction. Therefore, to enhance the robustness of motion estimation, AVS-EEM v8.1 proposes an adaptive downsampling-based strategy. For the current frame $x_t$ and its reference frame $x_{t-1}$, the method first applies a $d$-fold spatial downsampling, producing $x_t^d$ and $x_{t-1}^d$.  These downsampled frames are then upsampled back to the original resolution, yielding $\tilde{x}_t$ and $\tilde{x}_{t-1}$. The PSNR metrics between the original pair $PSNR_o(x_t,x_{t-1})$ and the processed pair $PSNR_d(\tilde{x}_t,\tilde{x}_{t-1})$ are calculated. If $PSNR_d > PSNR_o + \theta$, where  $\theta$ is a predefined threshold, motion estimation is performed on the downsampled frames $x_t^d$ and $x_{t-1}^d$; otherwise, estimation proceeds on the original-resolution frames $x_t$ and $x_{t-1}$. The method requires no bitstream overhead, as the decision is made internally at the encoder using the predefined parameters $d$ and $\theta$, which are kept consistent across all videos in the common test conditions.

\begin{figure}[t]
  \centering
   \includegraphics[width=\linewidth]{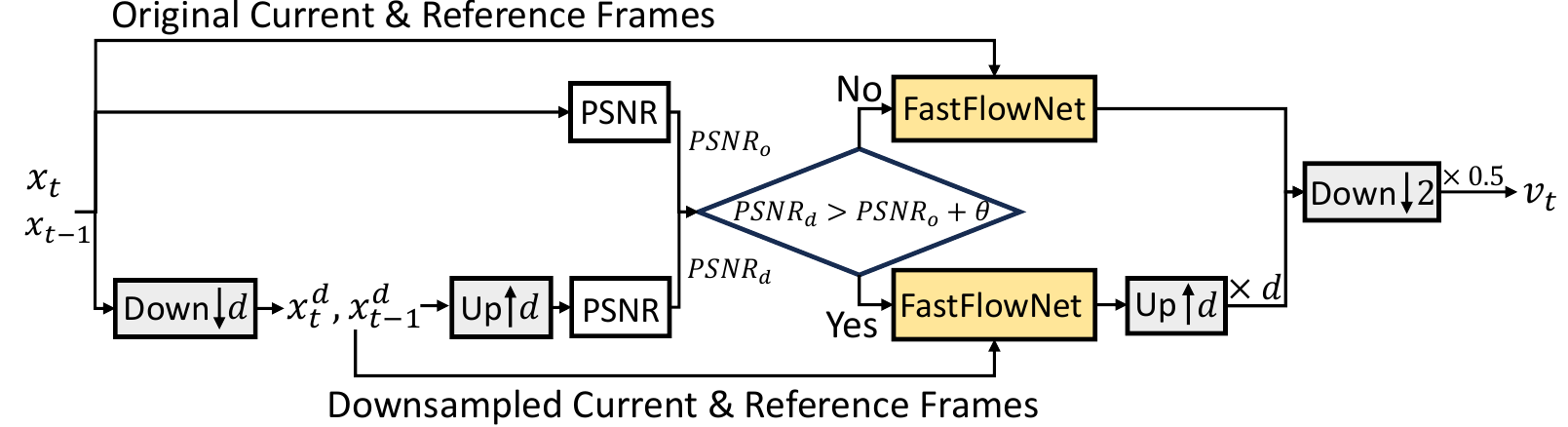}
      \caption{Adaptive downsampling-based motion estimation.}
   \label{fig:adaptive_motion_estimation}
\end{figure}

\subsubsection{Adaptive Skipping Threshold-based Entropy Coding}
In the motion and residual entropy models of AVS-EEM, the probability of the latent representations are assumed to follow Gaussian distributions. The entropy model estimates the mean and variance of the distributions to perform arithmetic coding. In some cases, directly encoding every sample of the latent representation may incur more bit costs. Skipping the entropy coding of certain samples and instead substituting them with the predicted mean may improve coding efficiency. Therefore, AVS-EEM v1.0, v2.0, and v6.1 successively propose and refine an adaptive skipping threshold-based entropy coding method. The decision to skip coding a sample is determined by comparing its estimated variance against a threshold $\eta_t$. A sample is skipped (i.e., replaced by the mean) if its variance is less than $\eta_t$; otherwise, it is entropy-coded. The threshold itself is adapted over time to account for error accumulation and the hierarchical quality structure. The threshold is defined as:
\begin{equation}
\eta_t=\beta_t \times e ^{(\frac{t}{N}-1) \times 0.3},
\end{equation}
 where $t$ is the frame index, $N$ is the total number of frames in the sequence, and $\beta_t$ is a quality-dependent coefficient. As encoding proceeds, reconstruction errors accumulate, leading to a gradual increase in $\eta_t$ , which encourages more skipping in later frames and thereby save bitrate costs.  To align with the hierarchical quality structure, $\beta_t$  is adjusted cyclically: a smaller  $\beta_t$ (yielding a lower $\eta_t$) is used for high-quality key frames, allowing finer detail preservation at the cost of moderate bit consumption; a larger $\beta_t$ (yielding a lower $\eta_t$) is applied for lower-quality non-key frames, promoting greater bit savings while maintaining acceptable reconstruction quality.

\subsubsection{Decoded Feature Refresh}
To exploit temporal correlation, AVS-EEM maintains a set of decoded features from the previous frame in a decoded buffer, which serve as priors for coding the subsequent frame. These include the motion-decoded feature $\hat{F}_{t-1}^v$, the motion temporal prior $\check{m}_{t-1}$, and the residual temporal prior $\hat{y}_{t-1}$. However, as encoding progresses, reconstruction errors accumulate over frames, causing the inaccuracies contained in these buffered features to increase.  To mitigate this drift, AVS-EEM v2.1, v6.1, and v9.2 successively introduce and refine a feature-refresh mechanism during inference: the stored features are periodically flushed, thereby truncating the accumulated error. The refresh period is set equal to the number of P-frames used in the cascaded-training stage, ensuring that the buffer is reset at intervals consistent with the temporal modeling capacity of the network.

\begin{figure*}[t]
  \centering
   \includegraphics[width=\linewidth]{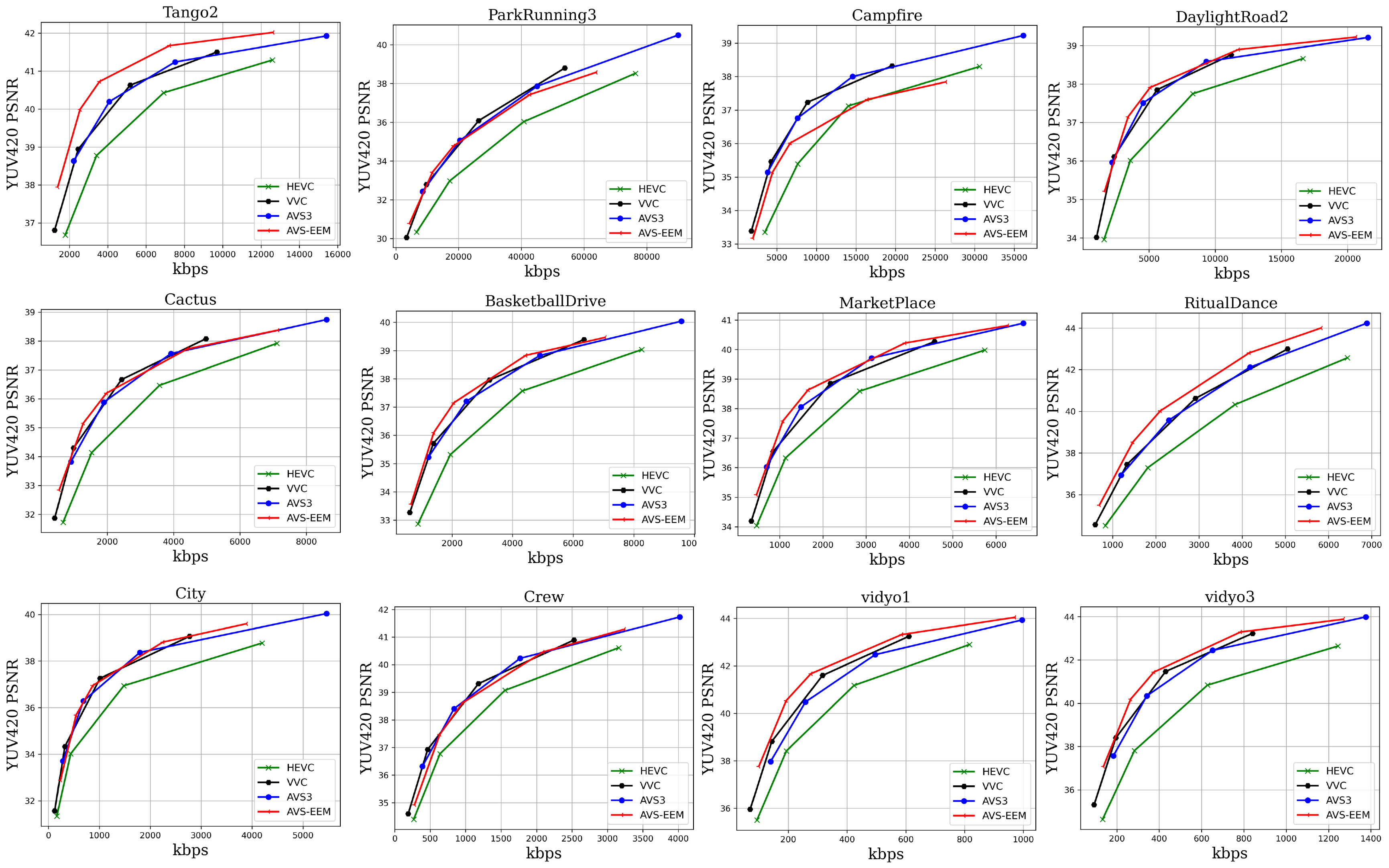}
      \caption{Rate-distortion curves of the latest AVS-EEM v9.2 model against the AVS3 reference software HPM-15.1, H.265/HEVC reference software HM-16.20, and H.266/VVC reference software VTM-13.2. The distortion is measured by a compound YUV-PSNR weighted in the 6:1:1 ratio (Y:U:V). }
   \label{fig:rd}
\end{figure*}

\section{Experiments}\label{sec:experiments}
\subsection{Experimental Setup}
\subsubsection{Training Dataset}
The training data of AVS-EEM are chosen based on the required sequence length. When the target training sequence length does not exceed 7 frames, the model is trained on the Vimeo-90k training set~\cite{xue2019video}, which contains 64610 clips of 7 frames each at $448 \times 256$ resolution. When the target length exceeds 7 frames, the BVI-DVC dataset~\cite{ma2021bvi} including 64940 clips of 64 frames cropped to $256 \times 256$ introduced in AVS-EEM v1.0, and 6920 clips of 30 frames at 360p resolution introduced in AVS-EEM v9.2 are used.
\subsubsection{Testing Dataset}
The evaluation of AVS-EEM is conducted on testing videos spanning three spatial resolutions: 4K (\emph{Tango2}, \emph{ParkRunning3}, \emph{Campfire}, \emph{DaylightRoad2}), 1080p (\emph{Cactus}, \emph{BasketballDrive}, \emph{MarketPlace}, \emph{RitualDance}), and 720p (\emph{City}, \emph{Crew}, \emph{vidyo1}, \emph{vidyo3}). All videos are in the YUV420 color format. Detailed specifications of these sequences---including resolution, frame rate, frame count, and bit depth---are provided in Table~\ref{table:testset}. 
\begin{table}[t]
\caption{Testing dataset of AVS-EEM.}
  \centering
\scalebox{0.9}{
\begin{tabular}{c|c|c|c|c}
\toprule[1.5pt]
Video           & Resolution & Frame Rate & Frame Count & Bit Depth \\ \hline
Tango2          & 3840$\times$2160  & 60         & 294          & 10        \\ \hline
ParkRunning3    & 3840$\times$2160  & 50         & 300          & 10        \\ \hline
Campfire        & 3840$\times$2160  & 30         & 300          & 10        \\ \hline
DaylightRoad2   & 3840$\times$2160  & 60         & 300          & 10        \\ \hline
Cactus          & 1920$\times$1080  & 50         & 250          & 8         \\ \hline
BasketballDrive & 1920$\times$1080  & 50         & 250          & 8         \\ \hline
MarketPlace     & 1920$\times$1080  & 60         & 300          & 10        \\ \hline
RitualDance     & 1920$\times$1080  & 60         & 300          & 10        \\ \hline
City            & 1280$\times$720   & 60         & 300          & 8         \\ \hline
Crew            & 1280$\times$720   & 60         & 300          & 8         \\ \hline
vidyo1          & 1280$\times$720   & 60         & 300          & 8         \\ \hline
vidyo3          & 1280$\times$720   & 60         & 300          & 8         \\
\bottomrule[1.5pt]
\end{tabular}}
\label{table:testset}
\end{table}
\begin{table}[t]
\caption{BD-rate performance of AVS-EEM v9.2 against the anchor HPM-15.1.}
  \centering
\begin{tabular}{c|c|c|c}
\toprule[1.5pt]
Video           & BD-rate-Y & BD-rate-U & BD-rate-V \\ \hline
Tango2          & --19.46\%  & --59.11\%  & --47.74\%  \\
ParkRunning3    & --6.09\%   & 44.22\%   & 46.77\%   \\
Campfire        & 9.31\%    & 177.75\%  & 31.04\%   \\
DaylightRoad2   & 1.90\%    & --48.40\%  & --19.13\%  \\ \hline
Cactus          & --0.73\%   & --35.73\%  & --38.22\%  \\
BasketballDrive & --4.98\%   & -43.78\%  & -34.92\%  \\
MarketPlace     & --1.31\%   & --47.05\%  & --46.93\%  \\
RitualDance     & --9.86\%   & --42.41\%  & --44.45\%  \\ \hline
City            & 1.11\%    & 26.75\%   & --37.37\%  \\
Crew            & 15.56\%   & -3.26\%   & --31.10\%  \\
vidyo1          & --18.84\%  & --52.21\%  & --48.02\%  \\
vidyo3          & --16.28\%  & --31.68\%  & --26.52\%  \\
\bottomrule[1.5pt]
4K              & --3.58\%   & 28.61\%   & 2.73\%    \\
1080p           & --4.22\%   & --42.24\%  & --41.13\%  \\
720p            & --4.61\%   & --15.10\%  & --35.75\%  \\ \hline
Average         & --4.14\%   & --9.58\%   & --24.72\%  \\
\bottomrule[1.5pt]
\end{tabular}
\label{table:rd}
\end{table}
\subsubsection{Test Configurations}
AVS-EEM is evaluated under a low-delay coding structure with the intra period set to --1, meaning that only the first frame is intra-coded and all following frames are inter-coded. This differs from MPAI-EEV~\cite{jia2023mpai}, which sets the intra period to 16 and encodes only a selected subset of frames. Additionally, while MPAI-EEV centrally crops input videos to multiples of 64, AVS-EEM spatially pads each video to a multiple of 16 before encoding. To align the rate points with the anchor, four Lagrange-multiplier $\lambda$ values are used: 4096, 2048, 512, 256, and 64. The I-frame is encoded with a YUV420-oriented I-frame model that shares the same architecture as the model in~\cite{sheng2022temporal} but is trained specifically for YUV420 PSNR optimization. Performance is measured in kbps for bit rate, and reconstruction quality is assessed by computing the PSNR of the Y, U, and V components separately. A compound YUV-PSNR weighted in the 6:1:1 ratio (Y:U:V) is reported as the primary quality metric. This differs from MPAI-EEV, which uses RGB PSNR to evaluate reconstruction quality. The anchor is the AVS3 reference software HPM-15.1 executed under its default common-test-condition (CTC) low-delay-P configuration. The performance difference between AVS-EEM and the anchor is quantified using the Bjøntegaard Delta (BD)-rate~\cite{bjontegaard2001calculation}.

\begin{table}[t]
\caption{BD-rate performance of all versions of AVS-EEM models against the anchor HPM-15.1.}
  \centering
\begin{tabular}{c|c|c|c}
\toprule[1.5pt]
Version & BD-rate-Y & BD-rate-U & BD-rate-V \\ \hline
v0.1    & 201.37\%  & 140.48\%  & 121.13\%  \\
v0.2    & 201.37\%  & 140.48\%  & 121.13\%  \\
v0.3    & 192.45\%  & 146.98\%  & 111.47\%  \\
v1.0    & 86.87\%   & 111.10\%  & 83.21\%   \\
v2.0    & 46.24\%   & 38.79\%   & 18.53\%   \\
v2.1    & 45.09\%   & 41.65\%   & 26.82\%   \\
v2.2    & 24.66\%   & 42.35\%   & 17.15\%   \\
v3.1    & 42.27\%   & 31.88\%   & 26.84\%   \\
v4.1    & 45.04\%   & 38.91\%   & 24.29\%   \\
v4.2    & 11.09\%   & 37.22\%   & 14.56\%   \\
v5.1    & 17.99\%   & 18.93\%   & 2.02\%    \\
v5.2    & 6.91\%    & 11.88\%   & -6.31\%   \\
v6.1    & 12.15\%   & 4.97\%    & -7.20\%   \\
v6.2    & 6.12\%    & 11.20\%   & -8.85\%   \\
v7.1    & 3.42\%    & --0.06\%   & --13.34\%  \\
v8.1    & --0.74\%   & --8.10\%   & --22.68\%  \\
v9.1    & --1.23\%   & --8.68\%   & --23.19\%  \\
v9.2    & --4.14\%   & --9.58\%   & --24.72\%  \\
\bottomrule[1.5pt]
\end{tabular}
\label{table:rd_all}
\end{table}

\subsection{Experimental Results}
\subsubsection{Compression Performance of the Latest Model}
The compression performance of the latest AVS-EEM v9.2 model is evaluated against the anchor HPM-15.1, with rate‑distortion curves illustrated in Fig.~\ref{fig:rd} and detailed BD‑rate results provided in Table~\ref{table:rd}. Across all test videos, AVS‑EEM v9.2 achieves an average BD-rate reduction of --4.14\% for the luma component, --9.58\% for chroma U, and --24.72\% for chroma V.
When examined by resolution, the model shows an average BD-rate of --3.58\% on luma, 28.61\% on chroma U, and 2.73\% on chroma V for 4K sequences. For 1080p sequences, the corresponding averages are --4.22\% on luma, --42.24\% on chroma U, and --41.13\% on chroma V. For 720p sequences, the averages are --4.61\% on luma, --15.10\% on chroma U, and --35.75\% on chroma V.
Significant gains are observed on several individual sequences. For instance, the 4K sequence Tango2 attains BD‑rate savings of --19.46\% in luma, --59.11\% in chroma U, and --47.74\% in chroma V. These results show that the average compression performance of AVS-EEM v9.2 surpasses that of the conventional AVS3 reference software HPM-15.1 across a diverse range of content types and resolutions. We also compare with the reference software of H.265/HEVC and H.266/VVC (HM-16.20 and VTM-13.2) under default low-delay-P configuration as complement.
\begin{table}[t]
\caption{BD-rate performance of main technologies of AVS-EEM.}
  \centering
\scalebox{0.9}{
\begin{tabular}{c|c|c|c}
\toprule[1.5pt]
Key Technologies                                                                                                                                 & BD-rate-Y & BD-rate-U                      & BD-rate-V                     \\ \hline
\begin{tabular}[c]{@{}c@{}}Content and Motion Feature-Assisted \\ Motion Compression\end{tabular}                                                & --15.90\% & --18.17\%                      & --17.56\%                     \\ \hline
\begin{tabular}[c]{@{}c@{}}Residual and Motion Temporal \\ Prior-Assisted Motion \\ Entropy Modeling\end{tabular}                                & --7.84\%  & --1.88\%                       & --5.14\%                      \\ \hline
\begin{tabular}[c]{@{}c@{}}Feature-Domain Group-Wise \\ Motion Alignment\end{tabular}                                                            & --3.61\%  & --3.59\%                       & --2.74\%                      \\ \hline
\begin{tabular}[c]{@{}c@{}}Reference Confidence-Assisted \\ Temporal Context Mining\end{tabular}                                                 & --10.20\% & --13.58\%                      & --10.73\%                     \\ \hline
\begin{tabular}[c]{@{}c@{}}Residual Confidence-Assisted \\ Frame Generator\end{tabular}                                                          & --11.11\% & --24.84\%                      & --26.13\%                     \\ \hline
\begin{tabular}[c]{@{}c@{}}Residual Temporal and Spatial \\ Prior-Assisted Residual Checkerboard \\ Autoregressive Entropy Modeling\end{tabular} & --3.89\%  & --3.57\%                       & --1.60\%                      \\ \hline
\begin{tabular}[c]{@{}c@{}}Group-Wise Multi-Frame\\  Cascaded Training\end{tabular}                                                              & --3.57\%  & \multicolumn{1}{l|}{--12.69\%} & \multicolumn{1}{l}{--13.66\%} \\ \hline
Hierarchical Quality-Based Training                                                                                                              & --13.71\% & --19.25\%                      & --10.93\%                     \\ \hline
Reference Quality-Aware Training                                                                                                                    & --9.31\%  & --15.99\%                      & --22.68\%                  \\ \hline
\begin{tabular}[c]{@{}c@{}}Adaptive Skipping Threshold-based \\ Entropy Coding\end{tabular}                                                      & --7.80\%  & 2.44\%                         & --1.06\%                      \\ \hline
\begin{tabular}[c]{@{}c@{}}Adaptive Downsampling-based\\  Motion Estimation\end{tabular}                                                         & --1.56\%  & --0.75\%                       & --1.16\%                      \\ \hline
Decoded Feature Refresh                                                                                                                          & --4.32\%  & --4.45\%                       & --3.78\%                      \\
\bottomrule[1.5pt]
\end{tabular}}
\label{table:performance_of_each_method}
\end{table}
\subsubsection{Performance Evolution Across Versions}
The BD-rate performance of all AVS-EEM versions against the anchor HPM-15.1 is summarized in Table~\ref{table:rd_all}. Aside from performance fluctuations in a few early versions due to codebase optimizations, each subsequent version has demonstrated notable performance gains. Starting from the initial version v0.1, which was significantly outperformed by HPM-15.1 (with BD-rates of 201.37\%, 140.48\%, and 121.13\% for Y, U, and V components, respectively), the latest version v9.2 achieves superior compression efficiency. This rapid progress highlights the continuous optimization of the framework. Beginning with versions v2.1 and v2.2, AVS-EEM introduced a distinction between short-trained (suffix ``.1") and long-trained (suffix ``.2") models. The short-trained models undergo a shorter training cycle and are used for rapid performance validation and integration of proposals during standard meetings. In contrast, the long-trained models are produced by the software maintainers after incorporating all adopted technical proposals and undergo extended training, typically yielding better compression performance. As shown in the table, for most version pairs (e.g., v4.1/v4.2, v5.1/v5.2), the long-trained model achieves more favorable BD-rate results than its short-trained counterpart, validating the positive impact of extended training on final performance.

\subsubsection{Individual Performance of Main Technologies}
The individual contributions of the core techniques integrated into AVS-EEM are quantitatively evaluated and summarized in Table~\ref{table:performance_of_each_method}. The reported BD-rate savings for each component are measured by comparing the full model against an ablated version where the corresponding technique is disabled, under the same common test conditions. This ablation analysis confirms the effectiveness of the proposed technologies. 

\begin{table}[t]
\caption{Computational complexities (MAC/pixel) of all modules of AVS-EEM v9.2.}
  \centering
\begin{tabular}{c|c}
\toprule[1.5pt]
Module                     & Complexity \\ \hline
Motion Estimation          & 27.44K     \\
Feature Extractor          & 18.43K    \\
Feature Extractor X        & 23.04K    \\
Feature Adaptor I          & 0.43K     \\
Feature Adaptor P          & 1.02K     \\
Feature Adaptor X          & 0.43K     \\
Motion Encoder             & 44.56K    \\
Motion Decoder             & 32.32K    \\
Motion Compensation        & 36.10K    \\
Motion Entropy Model Enc   & 12.93K    \\
Motion Entropy Model Dec   & 11.02K    \\
Residual Encoder           & 24.77K    \\
Residual Decoder           & 36.10K    \\
Residual Entropy Model Enc & 4.52K     \\
Residual Entropy Model Dec & 2.69K     \\
\toprule[1.5pt]
Encoder                   & 294.59K   \\
Decoder                   & 175.13K   \\
\bottomrule[1.5pt]
\end{tabular}
\label{table:complexity}
\end{table}

\begin{table}[t]
\caption{Computational complexities (MAC/pixel) of all versions of AVS-EEM.}
  \centering
\begin{tabular}{c|c|c}
\toprule[1.5pt]
Version & Encoder Complexity & Decoder Complexity \\ \hline
v0.1    & 281.48K  & 236.91K  \\
v0.2    & 281.48K  & 236.91K  \\
v0.3    & 281.48K  & 236.91K  \\
v1.0    & 278.41K  & 194.97K  \\
v2.0    & 364.55K  & 246.31K   \\
v2.1    & 364.55K  & 246.31K   \\
v2.2    & 364.55K  & 246.31K   \\
v3.1    & 270.99K  & 179.55K   \\
v4.1    & 273.30K  & 181.28K   \\
v4.2    & 273.30K  & 181.28K   \\
v5.1    & 289.15K  & 196.93K   \\
v5.2    & 289.15K  & 196.93K   \\
v6.1    & 275.56K  & 183.55K    \\
v6.2    & 275.56K  & 183.55K   \\
v7.1    & 267.15K  & 175.13K  \\
v8.1    & 294.59K  & 175.13K \\
v9.1    & 294.59K  & 175.13K \\
v9.2    & 294.59K  & 175.13K \\
\bottomrule[1.5pt]
\end{tabular}
\label{table:complexity_all}
\end{table}
\subsubsection{Computational Complexity}
The computational complexity of the latest AVS-EEM v9.2 is analyzed in terms of multiply-accumulate operations per pixel (MAC/pixel). Table~\ref{table:complexity} reports the complexities of all modules in the encoding and decoding processes. The total encoding complexity is 294.59 KMAC/pixel, and the total decoding complexity is 175.13 KMAC/pixel. Compared to MPAI-EEV, which has a complexity of 3127 KMAC/pixel, AVS-EEM is significantly more lightweight, making it more suitable for practical deployment. We also report the complexity of each version of AVS-EEM in Table~\ref{table:complexity_all}. Since the optical flow estimation network was frozen before the v8.1 version of the model, its complexity was not taken into account before that version.

\section{Future Work}\label{sec:future_work}
While the latest AVS-EEM model shows superior compression efficiency compared to the conventional AVS3 anchor, further advancements are essential to solidify its position as a next-generation video coding solution. Future research will focus on several key directions to bridge the remaining performance gaps and address practical deployment requirements.

\subsection{Intelligent Video Coding with Higher Compression Performance and Lower Computational Complexity}
Advancing AVS-EEM involves a dual pursuit: achieving higher compression performance while further reducing computational complexity. To enhance compression, future work will focus on improving long-term temporal modeling by scaling up cascaded training sequences, enriching the quality and diversity of training datasets, and developing more efficient neural architectures and coding tools. Concurrently, to lower complexity for broader deployment, efforts will prioritize variable bitrate training, architectural refinements, and model compression techniques, and will advance toward practical fixed-point quantization to ensure hardware-friendly, deterministic decoding. The synergy of these advancements is key to evolving AVS-EEM into a high-performance, low-complexity video coding standard.


\subsection{Intelligent Video Coding For Random-Access Scenario}
The current design of AVS-EEM is built around a P-frame coding structure that utilizes only past reference frames, making it optimal for low-delay applications. To mature into a comprehensive video coding standard for broader use cases---such as streaming and broadcasting---support for random-access configurations becomes essential. Random-access coding, typically realized through bi-directionally predicted (B-frame) structures, achieves higher compression efficiency by exploiting both past and future references, which is critical for high-efficiency storage and flexible seeking during playback. In response to this need, AVS-EEM has initiated exploratory work to extend its framework. Preliminary proposals~\cite{sheng2025bi} enabling neural random-access coding have been submitted, and exploratory experiments are now actively underway.

\subsection{Intelligent Video Coding for Perceptual Optimization}
Current video coding standards primarily target medium-to-high bitrate scenarios, often resulting in unsatisfactory reconstruction quality at lower bitrates. To address this limitation, leveraging advances in generative models presents a promising frontier for perceptual coding at low bitrates, where conventional methods struggle. AVS-EEM has initiated exploration in this direction, having received several proposals on generative coding. Key technical avenues include the use of generative adversarial network (GAN)~\cite{goodfellow2020generative} for model optimization, and the integration of more powerful video diffusion models~\cite{mao2025generative,chen2025generative} as priors for high perceptual quality generation. Exploratory experiments based on these concepts are currently underway, aiming to establish a new performance frontier for perceptually optimized, low-bitrate intelligent video coding.

\section{Conclusion}\label{sec:conclusion}
This paper presents a comprehensive overview of the standardization progress, key technologies, and experimental evaluation of the AVS End-to-End Intelligent Video Coding Exploration Model. We detail the model's development history and systematically introduce its core technical framework, which includes key model architectures,  training methodologies, and inference optimizations. Through sustained collaborative effort and iterative refinement, AVS-EEM has demonstrated a clear trajectory of rapid compression performance improvement, with its latest model achieving superior  compression efficiency over the conventional AVS3 reference software under stringent complexity and testing constraints. While this marks a significant milestone, future work is required to advance performance further, reduce complexity for broader deployment, and extend the model's applicability to variable-rate and random-access scenarios. The development of AVS-EEM represents a concrete and promising step toward the realization of a practical, next-generation intelligent video coding standard.

\section{Acknowledgment}\label{sec:Acknowledgment}
We would like to express our gratitude to Yunying Ge, Yibo Shi, Xinyao Chen, Yongqi Zhai, Yunlin Huang, Yihui Feng,~Jing Wang, Yin Zhao, Yucheng Sun, Yunzhuo Liu, Yuting Hu, Zongmiao Ye, Shiying Yin, Fangdong Chen, Xiaodi Shi, Dong Jiang, Jucai Lin, Shuhong Liao, Feng Ye, Tiange Zhang, Ziqing Ge, Xinyu Hang, Dongjian Yang, Kexiang Feng, Wenhong Duan, Xuan Deng, Hao Qi, Qi Zhang, Yanchen Zhao, Wenxuan He, Zhimeng Huang, Jiaqi Zhang, Xiandong Meng, Chuanmin Jia, Xiaopeng Fan, Debin Zhao, Siwei Ma, Shiqi Wang,~Renjie Zou, Yuntian Jiang, Yanbo Gao, Shuai Li, Jianjun Lei, Bin Li, Fang Xing, Pengfei Han, Haiqiang Wang,~Qi Wang, Lin Li, LiLi Chen, and XiaoXiao Zhou for their efforts of AVS-EEM project,

\bibliographystyle{ieeetr}
\bibliography{ref}
\begin{IEEEbiography}[{\includegraphics[width=1in,height=1.25in,clip,keepaspectratio]{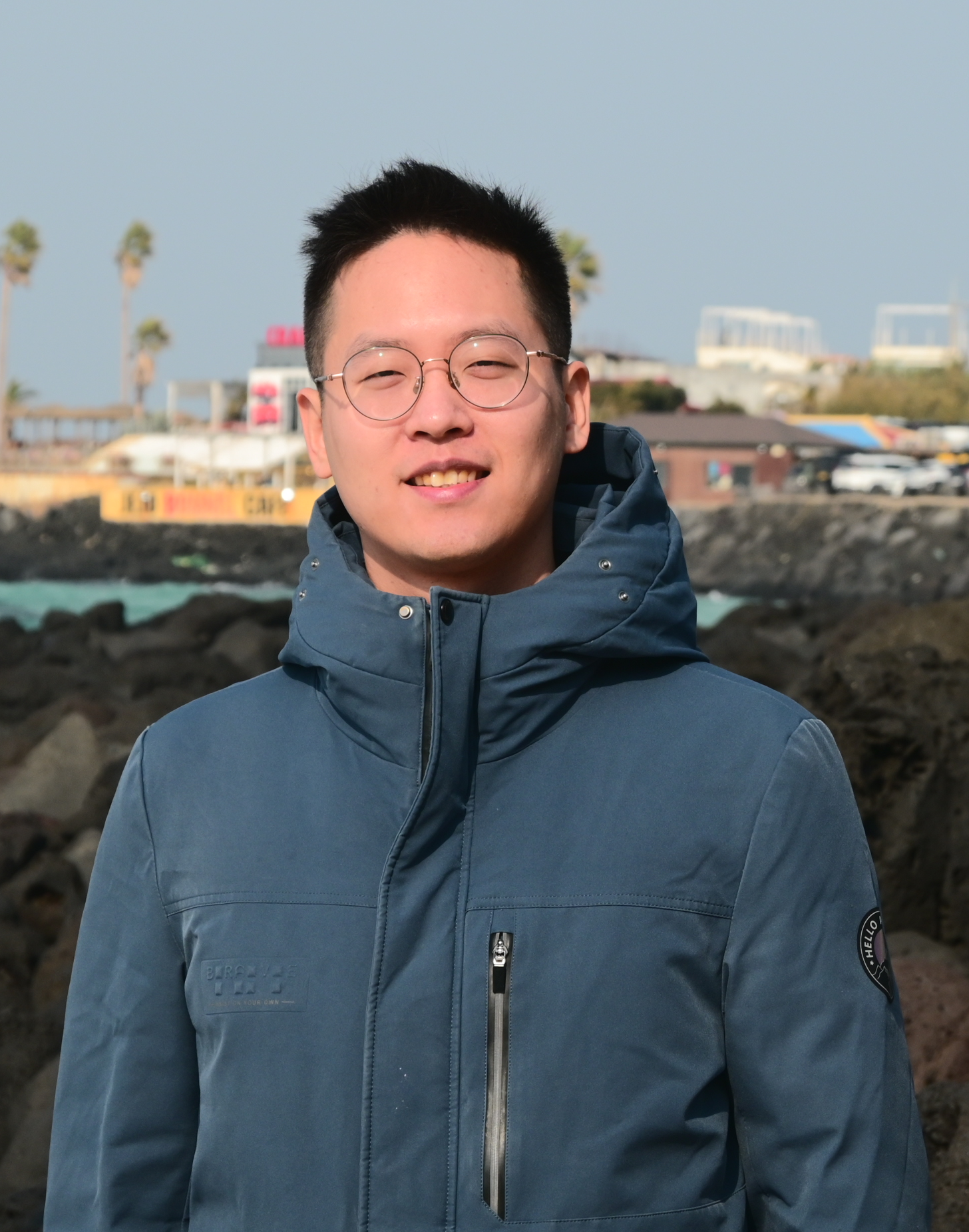}}]{Xihua Sheng} (Member, IEEE) received the B.S. degree in automation from Northeastern University, Shenyang, China, in 2019, and the Ph.D. degree in electronic engineering from University of Science and Technology of China (USTC), Hefei, Anhui, China, in 2024. 
He is currently a Postdoctoral Fellow in computer science from City University of Hong Kong (CityU). 
His research interests include image/video/point cloud coding, signal processing, and machine learning.
\end{IEEEbiography}
\begin{IEEEbiography}[{\includegraphics[width=1in,height=1.25in,clip,keepaspectratio]{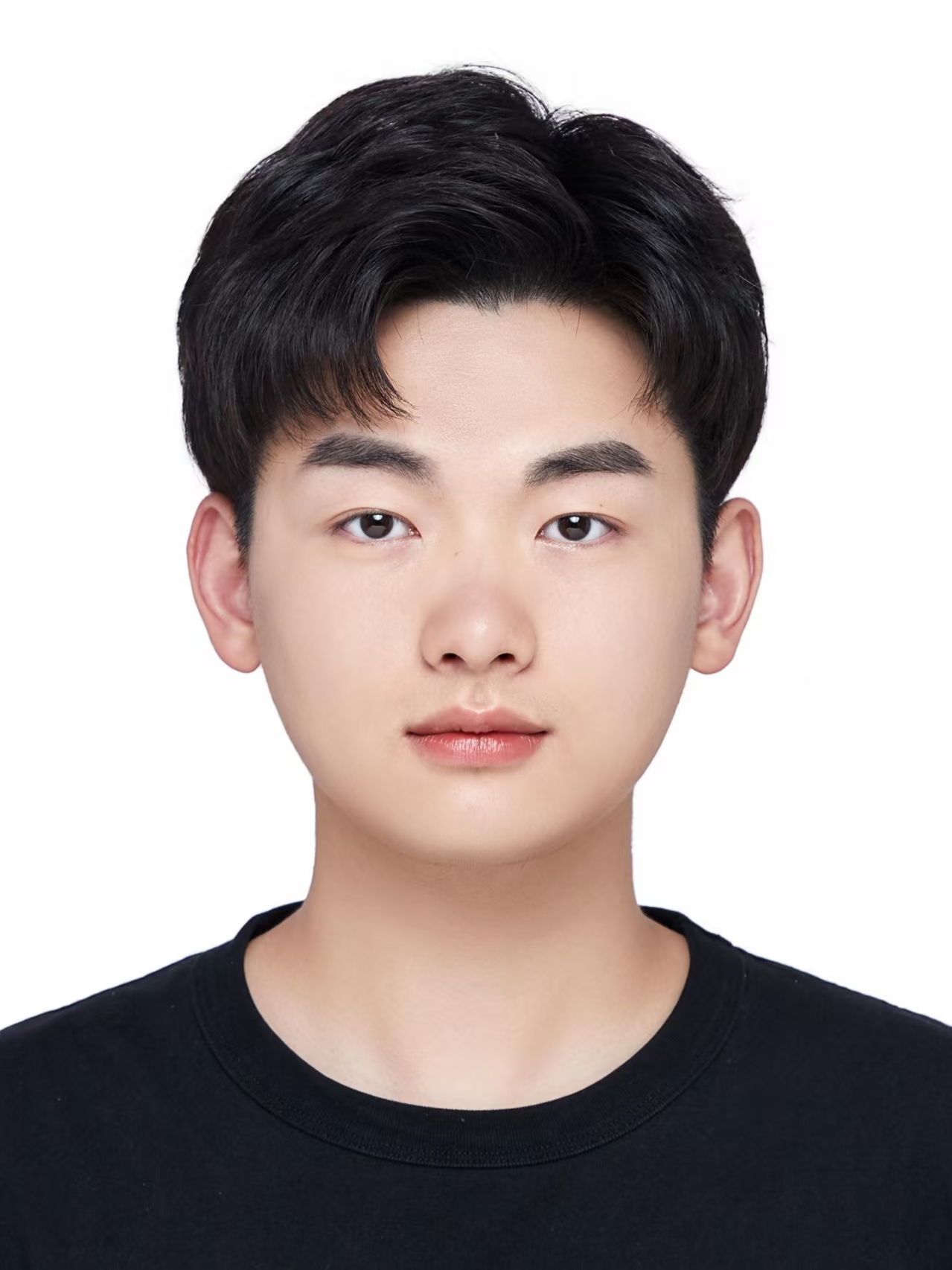}}]{Xiongzhuang Liang} received the B.S. degree from the School of Computer Science, China University of Geosciences, Wuhan, China, in 2023. He is currently pursuing the Ph.D. degree with the School of Artificial Intelligence and Data Science, University of Science and Technology of China, Hefei, China, under the supervision of Prof. Dong Liu.
\end{IEEEbiography}
\begin{IEEEbiography}[{\includegraphics[width=1in,height=1.25in,clip,keepaspectratio]{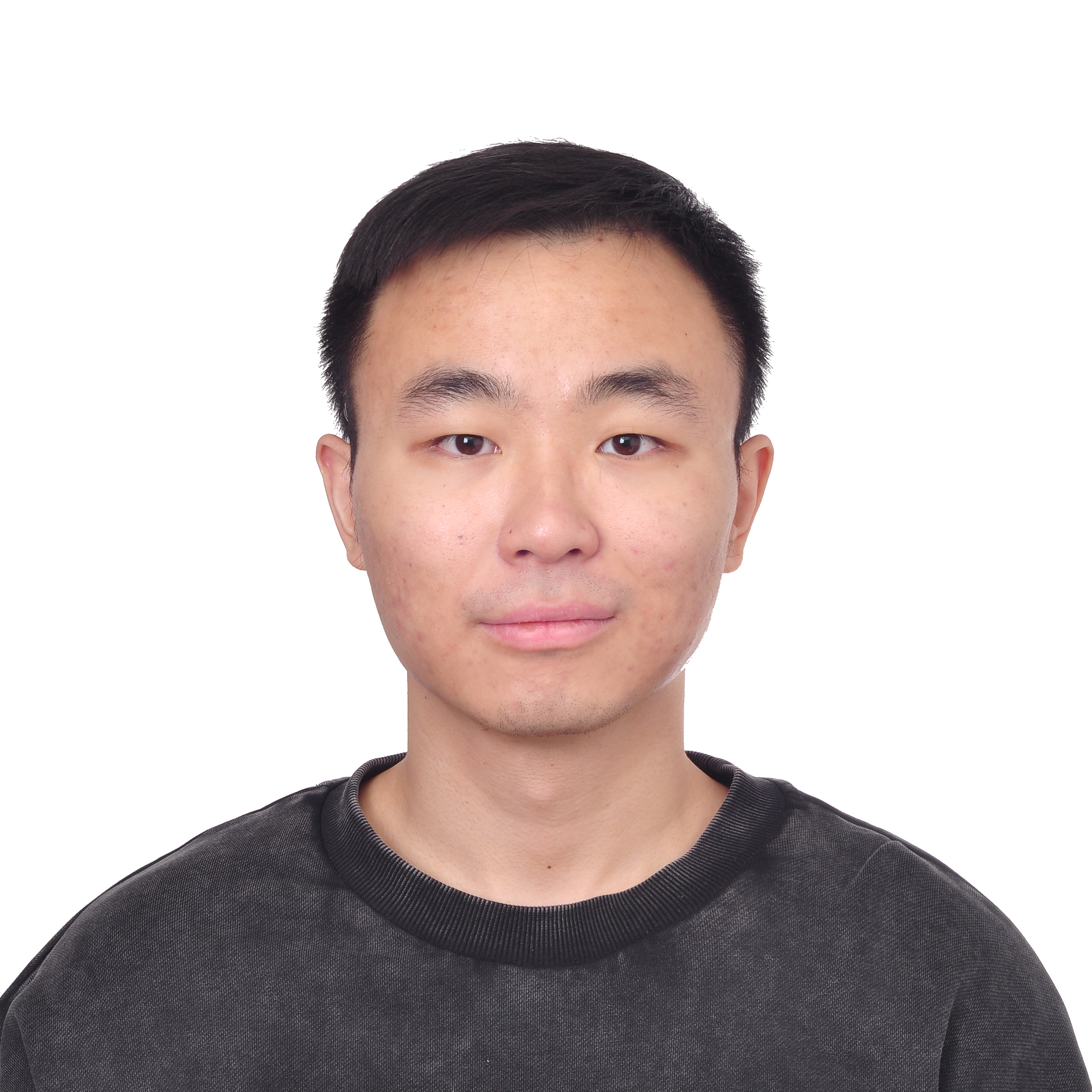}}]{ChuanboTang} received the B.S. degree in electronic information engineering from Hefei University of Technology, Hefei, China, in 2021. He is currently pursuing the Ph.D. degree in the Department of Electronic Engineering and Information Science at the University of Science and Technology of China, Hefei, China. His research interests include image/video coding, signal processing, and machine learning. 
\end{IEEEbiography}
\begin{IEEEbiography}[{\includegraphics[width=1in,height=1.25in,clip,keepaspectratio]{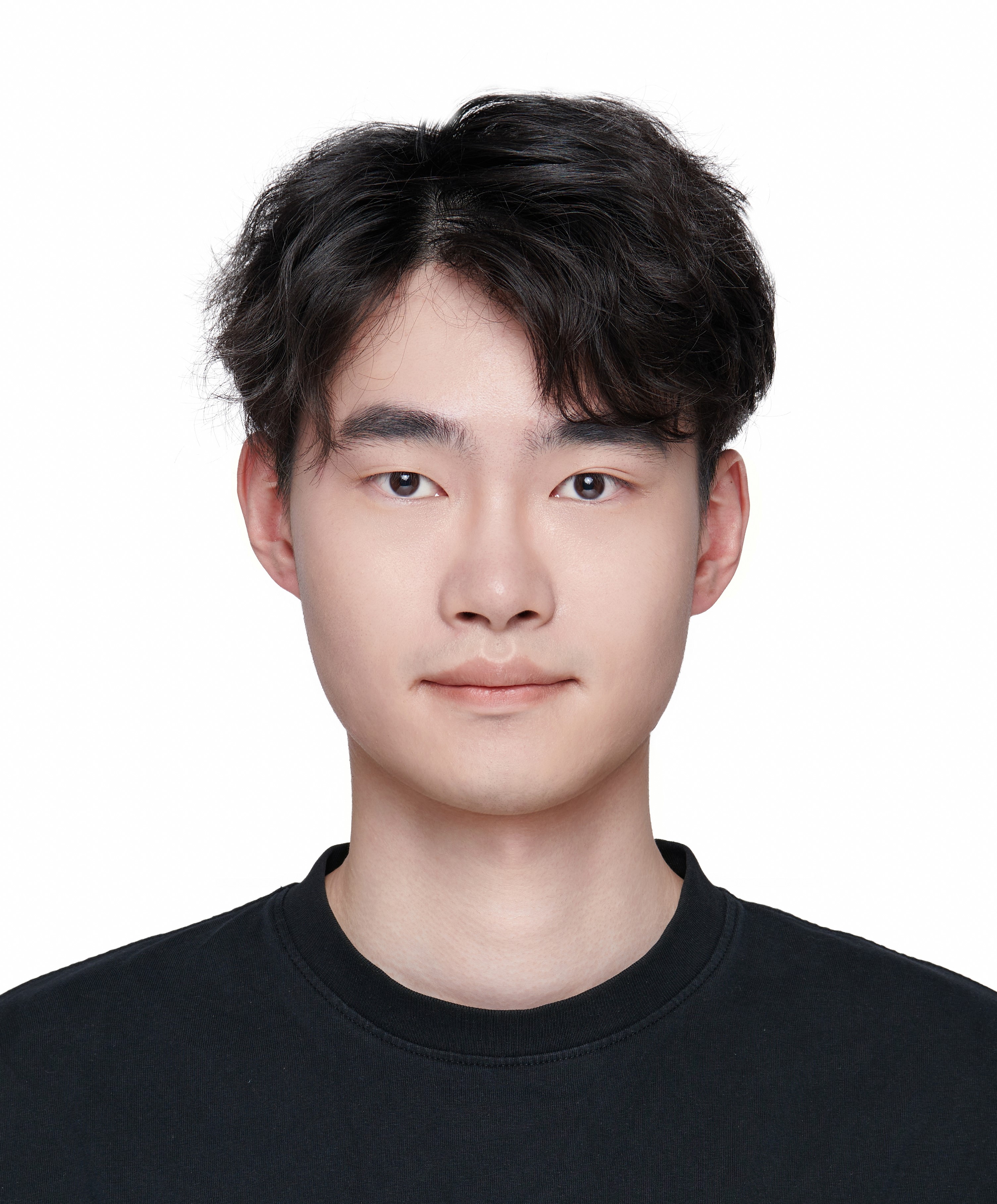}}]{Zhirui Zuo} received the B.S. degree in the School of Artificial Intelligence, Nanjing University, Nanjing, China, in 2023. He is currently pursuing the M.S. degree in the the Department of Electronic Engineering and Information Science at the University of Science and Technology of China, Hefei, China. His research interests include neural video compression.
\end{IEEEbiography}
\begin{IEEEbiography}[{\includegraphics[width=1in,height=1.25in,clip,keepaspectratio]{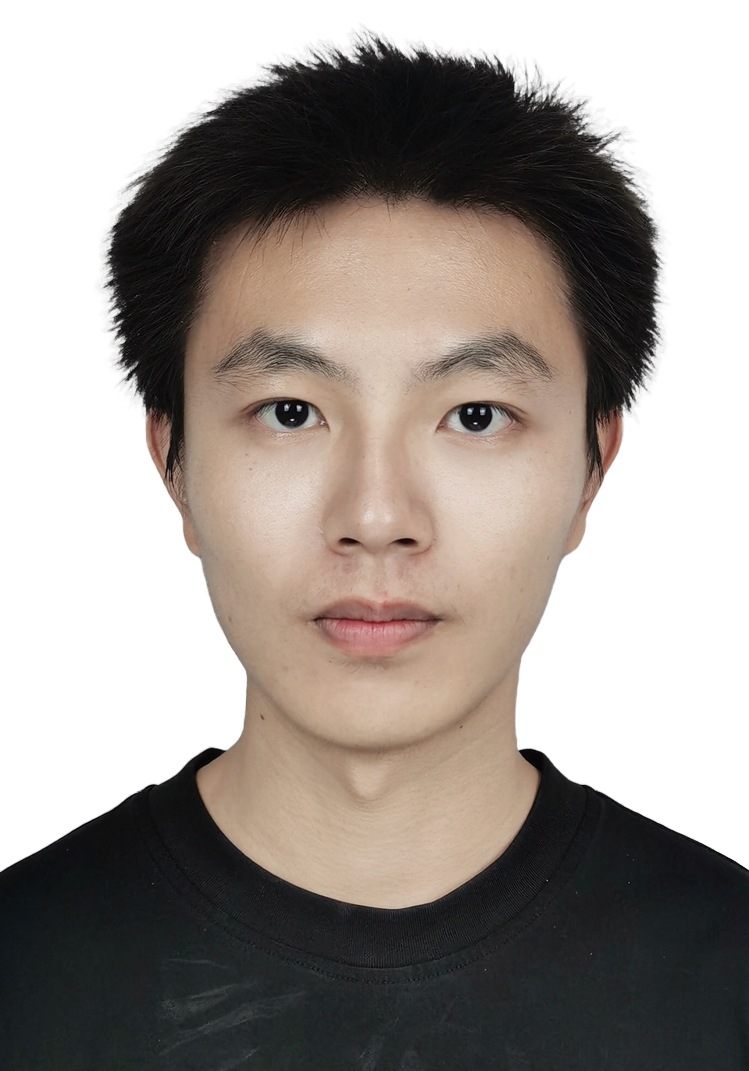}}]{YifanBian} received the B.S. degree in School of Electronic and Information Engineering from Tongji University, Shanghai, China in 2022. 
Currently, he is working toward the Ph.D. degree in the Department of Electronic Engineering and Information Science at the University of Science and Technology of China, Hefei, China.
His research interests include image and video processing, coding, and analysis.
\end{IEEEbiography}
\begin{IEEEbiography}[{\includegraphics[width=1in,height=1.25in,clip,keepaspectratio]{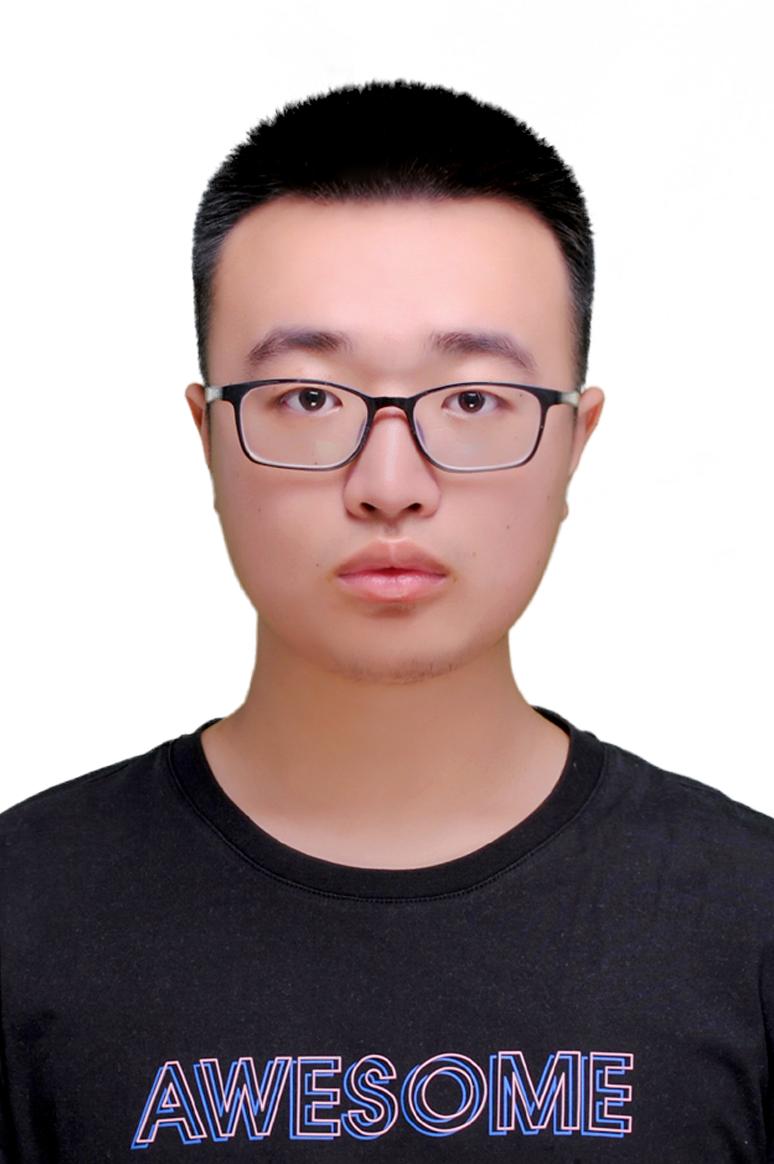}}]{Yutao Xie}
received the B.S. degree in artificial intelligence from Xidian University, Xian, China, in 2023. He is currently pursuing the Ph.D. degree in the Department of Electronic Engineering and Information Science at the University of Science and Technology of China, Hefei, China. His research interests include image and video coding.
\end{IEEEbiography}

\begin{IEEEbiography}[{\includegraphics[width=1in,height=1.25in,clip,keepaspectratio]{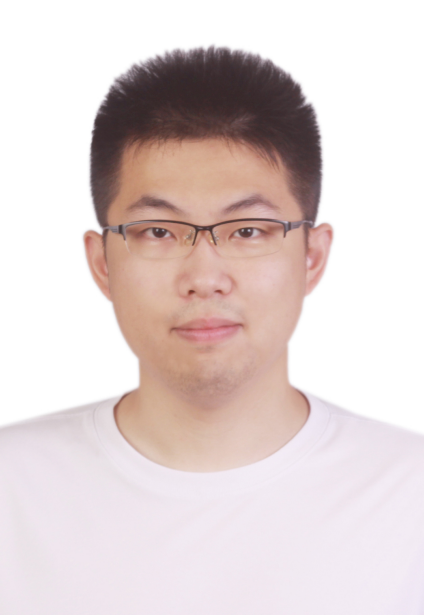}}]{Zhuoyuan Li}
(Member, IEEE) received the B.S. degree in communication engineering from Southwest Jiaotong University, China, in 2020, and the Ph.D. degree in electronic engineering and information science from the University of Science and Technology of China (USTC), China, in 2025. He will join The Hong Kong Polytechnic University (PolyU) as a Postdoctoral Fellow in 2026. His research interests include image and video coding. He has published over 10 papers in international journals and conferences. He won several technical challenges at ICIP 2024, MMSP 2024, and VCIP 2025. He received the PRCV 2025 Grand Challenge Outstanding Exploration Award and the NeurIPS 2025 Top Reviewer Award. He has submitted several technical proposals to ISO/IEC and AVS standardization activities.
\end{IEEEbiography}
\begin{IEEEbiography}[{\includegraphics[width=1in,height=1.25in,clip,keepaspectratio]{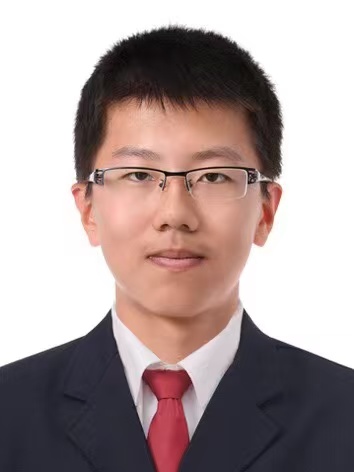}}]{Yuqi Li} received the B.S. degree in electronic information engineering from the University of Science and Technology of China (USTC), Hefei, China, in 2023. He is currently pursuing the Ph.D. degree in the Department of Electronic Engineering and Information Science at USTC. His research interests include image/video coding, image processing, and machine learning.
\end{IEEEbiography}
\begin{IEEEbiography}[{\includegraphics[width=1in,height=1.25in,clip,keepaspectratio]{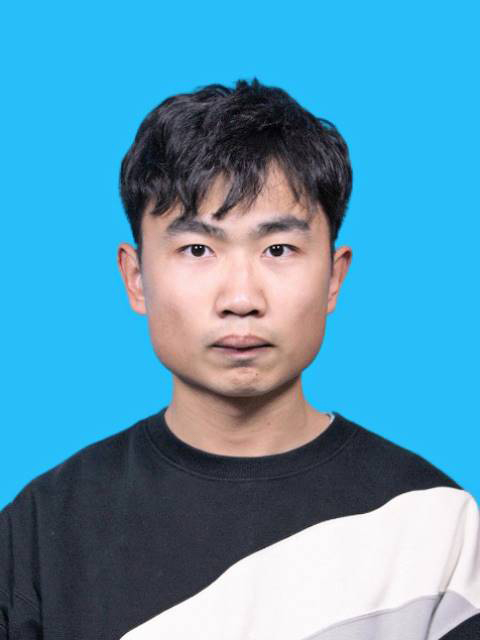}}]{Hui Xiang} received the B.S. degree in software engineering from Northeastern University, Shenyang, China, in 2024. He is currently pursuing the M.S. degree in the Department of Electronic Engineering and Information Science at the University of Science and Technology of China (USTC), Hefei, China. His research interests include image/video coding, machine learning, and computer vision.
\end{IEEEbiography}

\begin{IEEEbiography}[{\includegraphics[width=1in,height=1.25in,clip,keepaspectratio]{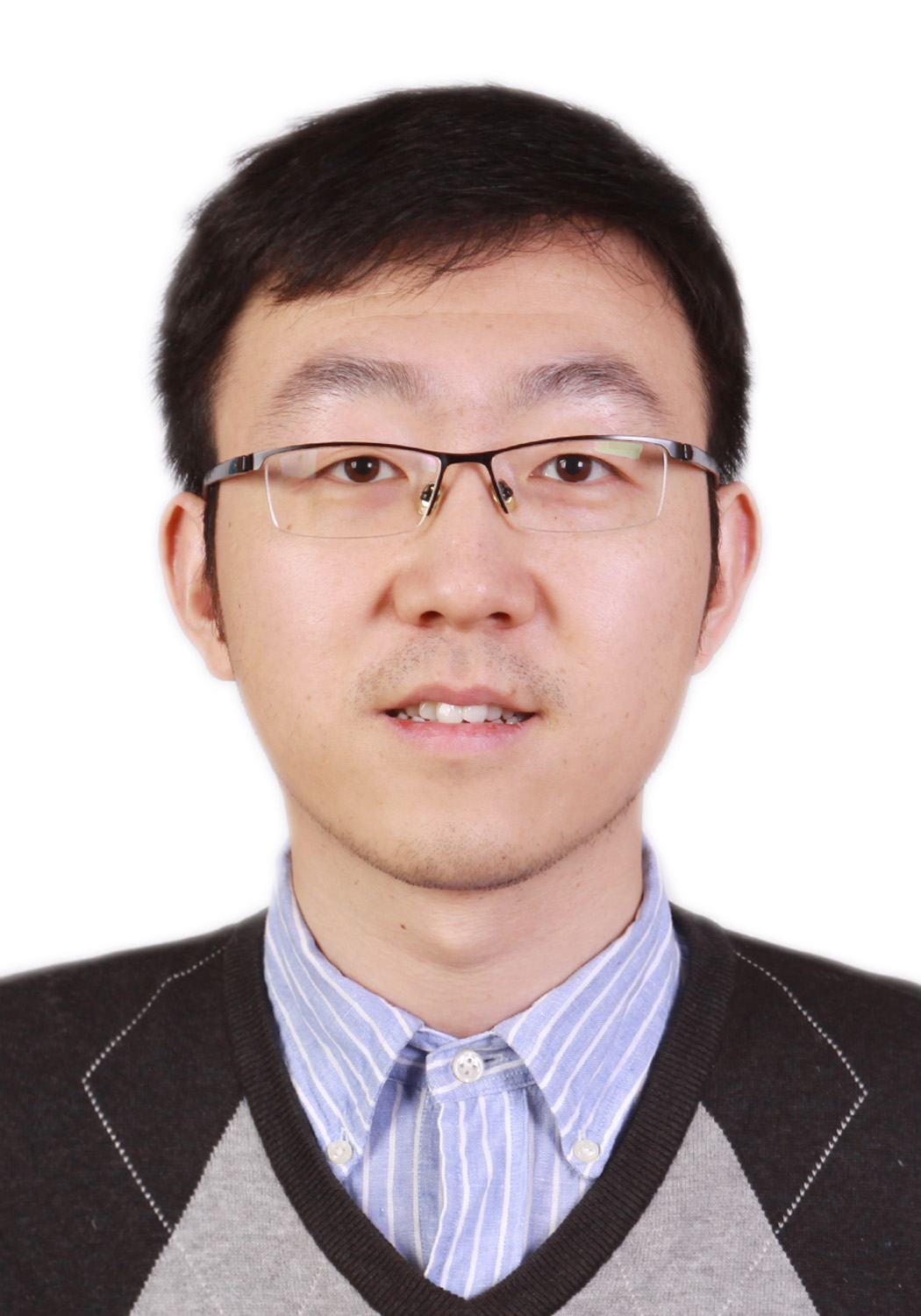}}]{Dong Liu}
(Senior Member, IEEE) received the B.S. and Ph.D. degrees in electrical engineering from the University of Science and Technology of China (USTC), Hefei, China, in 2004 and 2009, respectively. He was a Member of Research Staff with Nokia Research Center, Beijing, China, from 2009 to 2012. He has been a faculty member at USTC since 2012 and currently holds the position of full professor.

His research interests include image and video coding, processing, and visual intelligence. He has authored or co-authored more than 300 papers in international journals and conferences. He has more than 40 granted patents, and dozens of technique proposals adopted by standardization groups. He received 2009 IEEE TCSVT Best Paper Award and ISCAS 2025 Grand Challenge Top Creativity Paper Award. He is an elected member of IVMSP-TC of IEEE SP Society and an elected member of MSA-TC of IEEE CAS Society. He serves or had served as the Chair of IEEE 1857.11 Standard Sub-Working-Group, an Associate Editor for \textsc{IEEE Transactions on Image Processing}, a Guest Editor for \textsc{IEEE Transactions on Circuits and Systems for Video Technology}, and a TPC co-chair for ICME 2026.
\end{IEEEbiography}

\begin{IEEEbiography}[{\includegraphics[width=1in,height=1.25in,clip,keepaspectratio]{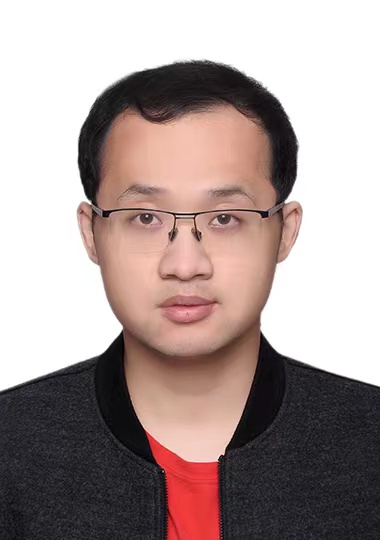}}] {Li Li} (M'17-SM'25) received the B.S. and Ph.D. degrees in electronic engineering from University of Science and Technology of China (USTC), Hefei, Anhui, China, in 2011 and 2016, respectively.
He was a visiting assistant professor in University of Missouri-Kansas City from 2016 to 2020.
He joined the department of electronic engineering and information science of USTC as a research fellow in 2020 and became a professor in 2022. \par
His research interests include image/video/point cloud coding and processing.
He has authored or co-authored more than 100 papers in international journals and conferences. 
He has more than 20 granted patents. 
He has several technique proposals adopted by standardization groups.
He received the Multimedia Rising Star 2023.
He received the Best 10\% Paper Award at the 2016 IEEE Visual Communications and Image Processing (VCIP) and the 2019 IEEE International Conference on Image Processing (ICIP).
He serves as a senior associate editor for \textsc{IEEE Transactions on Circuits and Systems for Video Technology} since 2026 and an associate editor for \textsc{IEEE Transactions on Multimedia} since 2025. 
\end{IEEEbiography}
\end{document}